\documentclass[lettersize,journal]{IEEEtran}
\usepackage{amsmath,amsfonts}
\usepackage{algorithmic}
\usepackage{algorithm}
\usepackage{array}
\usepackage[caption=false,font=normalsize,labelfont=sf,textfont=sf]{subfig}
\usepackage{textcomp}
\usepackage{stfloats}
\usepackage{url}
\usepackage{verbatim}
\usepackage{graphicx}
\usepackage{cite}
\usepackage{hyperref}

\usepackage[dvipsnames,table,xcdraw,svgnames]{xcolor}
\usepackage{paralist}
\usepackage{amsmath}
\usepackage{stfloats}

\usepackage{array}
\usepackage{footnote}
\usepackage{booktabs}
\makesavenoteenv{tabular}
\usepackage{multirow}
\usepackage{soul}
\usepackage{color}

\definecolor{myBlue}{HTML}{8AC5FF}
\definecolor{myOrange}{HTML}{F68511}
\definecolor{myGreen}{HTML}{22A316}
\definecolor{myRed}{HTML}{C22121}
\definecolor{myBlack}{HTML}{555555}

\definecolor{lightBlue}{HTML}{bceaf7}
\definecolor{lightGreen}{HTML}{e3f6bc}
\definecolor{lightRed}{HTML}{ffdede}

\definecolor{myBlack}{rgb}{0.0, 0.0, 0.0}

\begin{document}

\title{Volume-Based Space-Time Cube for\\Large-Scale Continuous Spatial Time Series}

\author{Zikun Deng, Jiabao Huang, Chenxi Ruan, Jialing Li, Shaowu Gao, and Yi Cai
\thanks{The work was supported by the National Natural Science Foundation of China (62402184) and National Key R\&D Program of China (2023YFB3309100). Y. Cai is the corresponding author.}
\thanks{Z. Deng, J. Huang, C. Ruan, and Y. Cai are with the School of Software Engineering, South China University of Technology and Key Laboratory of Big Data and Intelligent Robot (South China University of Technology), Ministry of Education. Z. Deng is also with the State Key Laboratory of Subtropical Building and Urban Science. Email: zkdeng@scut.edu.cn; jiabaoh20@outlook.com; 202130390185@mail.scut.edu.cn; ycai@scut.edu.cn}
\thanks{J. Li is with the Institute of Psychiatry, Psychology \& Neuroscience, King's College London. Email: jialing.1.li@kcl.ac.uk.}
\thanks{S. Gao is with the Greater Bay Area National Center of Technology Innovation. Email: gaoshaowu@ncti-gba.cn.}
\thanks{Manuscript received April 19, 2021; revised August 16, 2021.}}

\markboth{Journal of \LaTeX\ Class Files,~Vol.~14, No.~8, August~2021}%
{Shell \MakeLowercase{\textit{et al.}}: A Sample Article Using IEEEtran.cls for IEEE Journals}


\maketitle

\begin{abstract}
Spatial time series visualization offers scientific research pathways and analytical decision-making tools across various spatiotemporal domains.
Despite many advanced methodologies, the seamless integration of temporal and spatial information remains a challenge.
The space-time cube (STC) stands out as a promising approach for the synergistic presentation of spatial and temporal information, with successful applications across various spatiotemporal datasets.
However, the STC is plagued by well-known issues such as visual occlusion and depth ambiguity, which are further exacerbated when dealing with large-scale spatial time series data.
In this study, we introduce a novel technical framework termed VolumeSTCube, designed for continuous spatiotemporal phenomena.
It first leverages the concept of the STC to transform discretely distributed spatial time series data into continuously volumetric data.
Subsequently, volume rendering and surface rendering techniques are employed to visualize the transformed volumetric data. 
Volume rendering is utilized to mitigate visual occlusion, while surface rendering provides pattern details by enhanced lighting information.
Lastly, we design interactions to facilitate the exploration and analysis from temporal, spatial, and spatiotemporal perspectives.
VolumeSTCube is evaluated through a computational experiment, a real-world case study with one expert, and a controlled user study with twelve non-experts, compared against a baseline from prior work, showing its superiority and effectiveness in large-scale spatial time series analysis.
\end{abstract}


\section{Introduction}

\IEEEPARstart{S}{patial} time series, collections of time series data associated with geographic locations, are prevalent in many domains, such as environmental science, urban informatics, and atmospheric sciences, offering valuable insights within these fields.
Hereafter, we denote spatial time series as ``ST series'' for short.
Visualization is an effective and popular means for analyzing ST series, as it can present heterogeneous spatial and temporal information through easily perceivable and interactive graphics.
With the advancement of sensor technology, the scale of ST series data has become increasingly large.
By \textit{large-scale}, we refer to datasets containing millions of records, following previous studies~\cite{DBLP:journals/tvcg/LinsKS13,DBLP:journals/tvcg/LiuXR19}.
For example, China's air quality ST series spans an entire year with thousands of timestamps and covers the entire country with hundreds of ST series~\cite{DBLP:journals/tvcg/DengWLBZSXW22}, resulting in millions of records.
Such a large-scale feature poses challenges to effective visualization.

The effective composition of spatial and temporal visualizations has long been the most challenging problem in the visualization of spatiotemporal data beyond ST series~\cite{DBLP:journals/tvcg/SunLQW17,DBLP:journals/cvm/DengWLTXW23}. 
Linked views require users to pay context-switching costs to browse the spatial and temporal visualizations separately.
In contrast, an integrated view tightly organizes the spatial and temporal visualizations to reduce such costs.
The space-time cube (STC)~\cite{DBLP:journals/tvcg/KristenssonDABGHMNS09} is a kind of integrated view in a three-dimensional space, in which the space naturally occupies two dimensions (i.e., x-y plane), and the time constitutes the third dimension (i.e., z-axis).
The position of the visual graphics can be directly related to both space and time dimensions.
It is easy for users to accept and learn, and thus, users can perceive time and space simultaneously~\cite{DBLP:journals/tap/KjellinPSL10}.
Moreover, in the STC, the evolution of spatiotemporal phenomena can be effectively conveyed through temporal narratives, as geographic-related information is visualized along a continuous timeline~\cite{DBLP:journals/ijgi/MayrW18}.

The STC works well for various datasets, such as spatiotemporal events~\cite{DBLP:conf/iv/GatalskyAA04,DBLP:journals/tgis/NakayaY10,doi:10.1080/15230406.2023.2264749} and trajectories~\cite{kraak2003space,DBLP:journals/tvcg/FilhoSN20}.
Yet, adopting the STC to visualize large-scale ST series is still in an exploratory stage.
Prior studies~\cite{DBLP:conf/iv/ThakurH10,Tominski2005STC,DwyerGallagher2004STC} adopted the STC to visualize ST series (e.g., \autoref{fig:baseline}), but only a few timestamps (e.g., $\le 30$) are applicable.
It remains unexplored how to enable in-depth visual analysis of large-scale, highly dynamic, and multi-correlated ST series data based on the STC.
To this end, this study aims to enhance the STC for visualizing large-scale ST series.
However, two major challenges are posed.

\textbf{Effective presentation of spatiotemporal patterns.}
Due to the characteristics of the STC, each reading value of the ST series can directly correspond to time and space.
However, the effective presentation of massive values is non-trivial.
With Thakur and Hanson's method~\cite{DBLP:conf/iv/ThakurH10}, the graphic plot of each ST series tends to occlude each other (\autoref{fig:baseline}).
Besides, the dataset's hidden spatiotemporal patterns (e.g., hotspots and propagation processes) cannot be revealed clearly.
New data transformation and visualization strategies are strongly required to summarize and present the large-scale ST series in less visual occlusion.

\textbf{Flexible exploration of spatial and temporal domains.}
The rendered STC comprising large-scale ST series will be too large for users to navigate the time dimension to identify periods of interest.
Also, in the huge cube, the visual graphics can be far away from the map, and it may be difficult for users to associate the graphics with the geographical context.
The depth channel also exacerbates the issue of spatial correspondence.
It is cumbersome to pinpoint the desired selection in three-dimensional space because the projection line of the object to the screen may intersect with multiple targets.
Therefore, flexible interactions should be designed and implemented to accommodate the STC environment with large-scale ST series.

This study focuses on ST series of phenomena that are continuous across both space and time, a characteristic typically found in natural phenomena, in contrast to other phenomena such as tourist visits to different attractions and demographic trends across countries.
We propose a visualization technique named VolumeSTCube that addresses the aforementioned challenges.
For the first challenge, we propose a transformation framework that transforms large-scale ST series into a volume visualization.
In particular, the framework adopts interpolation to convert discretely distributed values into continuously distributed volumetric data following the STC schema.
Afterward, the volumetric data is rendered with the volume and surface rendering techniques in a less cluttered manner.
For the second challenge, we design a set of flexible interactions to facilitate exploring the STC from temporal, spatial, and spatiotemporal perspectives.
The volume can be sliced and highlighted for spatial and temporal exploration, respectively.
A voxel cluster-based interaction builds upon the volume slicing and highlights to enable easy selection and investigation of spatiotemporal patterns.

VolumeSTCube is evaluated as follows.
First, we conducted a case study on a real-world dataset with VolumeSTCube, performed by one expert, to demonstrate its intuitiveness, usefulness, and effectiveness fully. 
Second, we performed a task-based controlled user study with twelve non-expert participants to compare VolumeSTCube with the prior STC-based ST series visualization~\cite{DBLP:conf/iv/ThakurH10}.
VolumeSTCube achieves better performance and receives positive user feedback regarding the visualization and interaction.

In sum, our contributions can be summarized as follows:
\begin{compactitem}
    \item A visualization technique named VolumeSTCube that incorporates a data transformation framework, volume visualization techniques, and tailored spatiotemporal interactions to visualize large-scale ST series in a space-time cube effectively.
    \item Comprehensive evaluations for VolumeSTCube with a computational experiment, a real-world case study, and a task-based controlled user study.
\end{compactitem}

\section{Related Work~\label{sec:relatedwork}}
This section presents the prior studies related to our study from three aspects, namely, spatiotemporal analysis, spatial time series visualization, and space-time cube.

\subsection{Spatiotemporal Analysis}
Spatiotemporal analyses are commonly seen in various domains.
Although many automated algorithms for spatiotemporal analysis have been proposed, the visual analysis that keeps humans in the loop is an important approach~\cite{DBLP:books/daglib/0015278}.
Below are some examples of ST series analysis, which is the focus of our study.
Deng et al.~\cite{DBLP:journals/tvcg/DengWXBZXCW22} developed a tailored dynamic graph visualization to elucidate the dynamic causal relationships within ST series, facilitating spatially and temporally aware interpretation and validation.
Li et al.~\cite{DBLP:journals/tvcg/LiCZAA19} designed a visualization framework that enables the interactive extraction and exploration of event co-occurrences across ST series.

Different from such kinds of specific tasks of spatiotemporal analysis, our study focuses on one of the basic tasks of analysts, obtaining the \textbf{temporal trends} of the spatiotemporal observations within \textbf{the spatial context}, which can be seen in various domains~\cite{yu2017analysing,xu2020spatiotemporal,muthoni2019long}.
For example, in Yu and He's traffic data analysis~\cite{yu2017analysing}, they learned that ``\textit{during the morning peak, the major departure locations are dispersed across Yuexiu district, Liwan district, and western Haizhu district $\dots$}''
In Muthoni et al.'s meteorology analysis~\cite{muthoni2019long}, they concluded that ``\textit{the annual rainfall anomalies revealed that ESA (Eastern and Southern Africa) region received above normal rainfall in 1982, 1989, 1997 and 2006, with the latter being the most severe.}''
However, the basic task becomes rather difficult given large-scale ST series that expand a vast spatial range, involve a long time period with thousands of timestamps, and exhibit inter-correlated and highly time-varying natures.

Our research problem is how to design a visualization for large-scale ST series that supports analyzing temporal trends of spatiotemporal observations within the spatial context.

\subsection{Spatial Time (ST) Series Visualization~\label{sec:related_stvis}}

ST series visualizations can be classified into linked views and integrated views based on the composition of spatial and temporal visualizations.

\textbf{Linked views.}
As implied by its name, linked views utilize multiple views to depict temporal and spatial information separately and link them through interactive visualization.
One of the views for spatial information can be a geographical map~\cite{DBLP:journals/tvcg/ZhangXTGLZ25} or a structure diagram~\cite{DBLP:journals/tvcg/LiuWTDXZYZW23}, and another view for temporal information can be a bar chart~\cite{DBLP:journals/tvcg/YangZFJCSS23}, an area chart~\cite{DBLP:journals/vi/ZhaoWWLWW22}, heat matrices~\cite{10290915}, and spatial neighborhood-preserving 1D timelines~\cite{DBLP:journals/cgf/FrankeMKK21,DBLP:journals/tvcg/KoppW23}.
Various strategies can be used to establish the link between views.
For example, Yang et al.'s method~\cite{DBLP:journals/tvcg/YangZFJCSS23} relied on user interactions: the temporal visualization at a location will be displayed after the user clicks the location on the map.
In contrast, Li et al.'s method~\cite{9552191} and Zhao et al.'s method~\cite{DBLP:journals/vi/ZhaoWWLWW22} adopted explicit encodings: each temporal visualization had a unique location ID that can be exactly related to the geographic context.

These methods are limited in scalability.
Users need to make many spatial and time selections to browse a large-scale ST series dataset that has wide spatial coverage and a large time span.
Pre-summarizing or mining patterns within ST series before visualization, rather than directly depicting ST series, can lead to greater scalability but may result in information loss and disrupt the temporal narrative.
Recently, Deng et al.~\cite{deng2023visualizing} extracted evolution patterns from the ST series and organized them into narrative-preserving Storyline layout, proposing a scalable visualization technique.

However, the main issue of linked views is that users need to frequently switch between multiple views to obtain temporal and spatial information, paying context switching costs and facing the cognitive burden~\cite{DBLP:conf/avi/BaldonadoWK00,convertino2003exploring}

\textbf{Integrated view.}
Integrated views are designed to tightly present spatial and temporal information, allowing users to analyze spatiotemporal data in one view without switching contexts.
For example, Sun et al.~\cite{DBLP:journals/tvcg/SunLQW17} embedded traffic time series visualizations into road segments on the map after the roads are topologically expanded.
Li et al.~\cite{DBLP:conf/ieeevast/LiZM14} designed a layout where timeline-based time series visualizations extend radially from the corresponding locations on the map.
The glyph is also useful for integrating heterogeneous information~\cite{DBLP:journals/tvcg/YingSDYTYW23,DBLP:journals/tvcg/YingTLSXYW22}. 
For example, in Deng et al.'s glyph~\cite{DBLP:journals/tvcg/DengWCLWBZW20}, the temporal occurrence of a propagation pattern was wrapped on a circular map where the pattern is displayed.
While this approach minimizes context switching costs, it employs distinct visual channels to encode temporal and spatial information separately, thereby limiting its intuitiveness.

The space-time cube (STC) can be considered as an integrated view.
The STC-based visualization of ST series is less explored compared to linked views.
Existing studies~\cite{DBLP:conf/iv/ThakurH10,Tominski2005STC,DwyerGallagher2004STC} primarily represent each ST series as a column within the space-time cube, positioning each column according to its geographical location.
While STC-based visualizations provide an integrated presentation of spatial and temporal information, they face scalability issues when dealing with large-scale ST series, as will be discussed in the next subsection.

Our study attempts to improve the integrated spatiotemporal visualization technique STC for large-scale ST series.

\subsection{Space-Time Cube}
The space-time cube (STC) was originally proposed by T.~Hägerstrand in the early 70s~\cite{ilagcrstrand1970people} to depict the life histories of humans.
The STC seamlessly integrates spatial and temporal dimensions within a 3D cube space, comprising a geographic map (i.e., a plane) and a vertical time axis.
Such a design enables users to view the entire spatiotemporal dataset in a single view with both spatial context and temporal information concurrently.
This sets it apart from 2D visualization techniques, which typically require users to navigate slider controls for temporal exploration or toggle between a geographic map and a separate time-oriented visualization.

\textbf{STC applications.}
The STC has been successfully applied in the visualization of spatiotemporal datasets in various domains, such as earthquake events~\cite{DBLP:conf/iv/GatalskyAA04}, crime clusters~\cite{DBLP:journals/tgis/NakayaY10}, mobilities~\cite{DBLP:conf/vr/FilhoSSN24}, and eye-tracking recordings~\cite{Kurzhals2013STCEye}.
However, the STC has the same well-known limitation as other 3D data representations: occlusion~\cite{DBLP:journals/tvcg/ElmqvistT08}.
For example, if there are more time series and more timestamps, there will be occlusion between columns (\autoref{fig:baseline}).
The user may suffer from the process of searching the patterns from many multiple columns that are separately distributed in the cube.

Many studies proposed to mitigate the occlusion problem through volume visualization and user interactions~\cite{DBLP:journals/tvcg/ElmqvistT08}.

\textbf{Volume visualization.}
In the earliest applications, the integration of volume data and space-time cubes (STC) was prevalent in video visualizations, as the sequence of frames in a video inherently constitutes volumetric data. Video visualizations aim to efficiently capture activities within videos~\cite{4530419,Daniel2003Video,Romero2008Video,DBLP:journals/tvcg/ChenBHWET06} and identify areas of interest for viewers~\cite{Bruder2019Gaze}.
These visualizations focus on summarizing the video without facilitating interactive exploration of the details, whereas ST series capture many local but important patterns.

Such a combination was also adopted in the geographic domain.
Demšar and Virrantaus~\cite{Virrantaus2010STCTrajetories} transformed trajectories in STC into volumetric density representations using 3D kernel density estimation (KDE).
Similarly, Nakaya and Yano~\cite{DBLP:journals/tgis/NakayaY10} represented crime clusters as volumetric densities in STC via 3D KDE.
The visualizations generated through volume rendering can mitigate occlusion issues by revealing the overall distribution characteristics of data in a 3D space.
These studies, however, have been conducted on small datasets, leaving the interactive exploratory approach for large-scale ST series unexplored.

\textbf{Interactions in STC.}
Elaborate user interactions, such as probing~\cite{DBLP:conf/vrst/Elmqvist05}, rotation~\cite{DBLP:journals/tvcg/YeCCLLLGW23}, and cutting~\cite{carpendale1999tardis} can be applied to ease visual occlusion and enhance usability via flexible 3D exploration.
Bach et al.~\cite{Bach2017STCInteraction} reviewed the existing STC interactions and presented a descriptive framework for interacting with a generalized space-time cube.
However, a few interactions were designed to explore STC with large-scale spatiotemporal data.
Filho et al.~\cite{DBLP:conf/vr/FilhoSSN24} introduced a set of interactions for selecting and filtering large-scale taxi trips in the STC from spatial and temporal perspectives.
Our study aims to propose STC interactions for ST series of spatiotemporal phenomena.
Specifically, we follow Bach et al.'s framework~\cite{Bach2017STCInteraction} and design STC interactions from temporal, spatial, and spatiotemporal perspectives to assist users in exploring large-scale ST series.

\vspace{3pt}
This study reassesses the suitability of the STC for large-scale ST series.
Ultimately, we propose effective data transformation and visualization techniques to enhance the presentation of large-scale ST series within the STC.

\section{Background}
This section describes the data formats used in this study, and introduces the research problem with a potential solution.

\subsection{Data Description}
In this study, an ST series dataset comprises $S$ ST series, i.e., $V_1$, $V_2$, $\dots$, $V_S$.
Each \textbf{ST series} $V_i$ is a series of values in chronological order, which is collected from a monitoring sensor (e.g., an air quality monitoring station) with a fixed geographic position.
Formally, $V_i = \{v_{i,1}, v_{i,2}, \dots, v_{i,T}\}$, where $T$ is the maximum timestamp and denotes the number of values in the ST series, and $v_{i,t} \in V_i$ denotes the reading value of the sensor at the timestamp $t$.
For example, \autoref{fig:pipeline}A comprises five ST series; $V_1$ is located at the top right corner of the geo-space and its values at the second and third timestamp are 50 and 100, respectively.

A \textbf{space-time volume} \(\mathcal{V}\) is a 3D scalar field or volumetric data. It comprises \(m \times n \times T\) samples \((s, t, v)\), where \(s = (x, y)\) is a 2D spatial index, \(t\) is a 1D temporal index, and \(v\) is a real value. 
\(1 \le x \le m\), \(1 \le y \le n\), and \(1 \le t \le T\).
Each sample \((s, t, v)\) represents the value \(v\) at a 3D location \((x, y, t)\), i.e., \(v\) is observed at the timestamp \(t\) in the geographic position \(s = (x, y)\).
In practical situations, each 3D location is represented by a small hexahedral cube called a voxel.
In particular, we use \(\mathcal{V}^{(x_0, y_0, \_)}\) to denote the ST series at the geographic position \((x_0, y_0)\) extracted from the space-time volume \(\mathcal{V}\).
Taking the space-time volume in \autoref{fig:pipeline}B as an example, there is a \(3 \times 3 \) grid (i.e., $m = n = 3$) in the geographic space and 3 timestamps (i.e., $T = 3$), resulting \(3 \times 3 \times 3 \) samples or voxels in total.
The voxels colored pink denote \(\mathcal{V}^{(1, 3, \_)}\).

\begin{figure}[tb]
  \centering
  \includegraphics[width=\columnwidth]{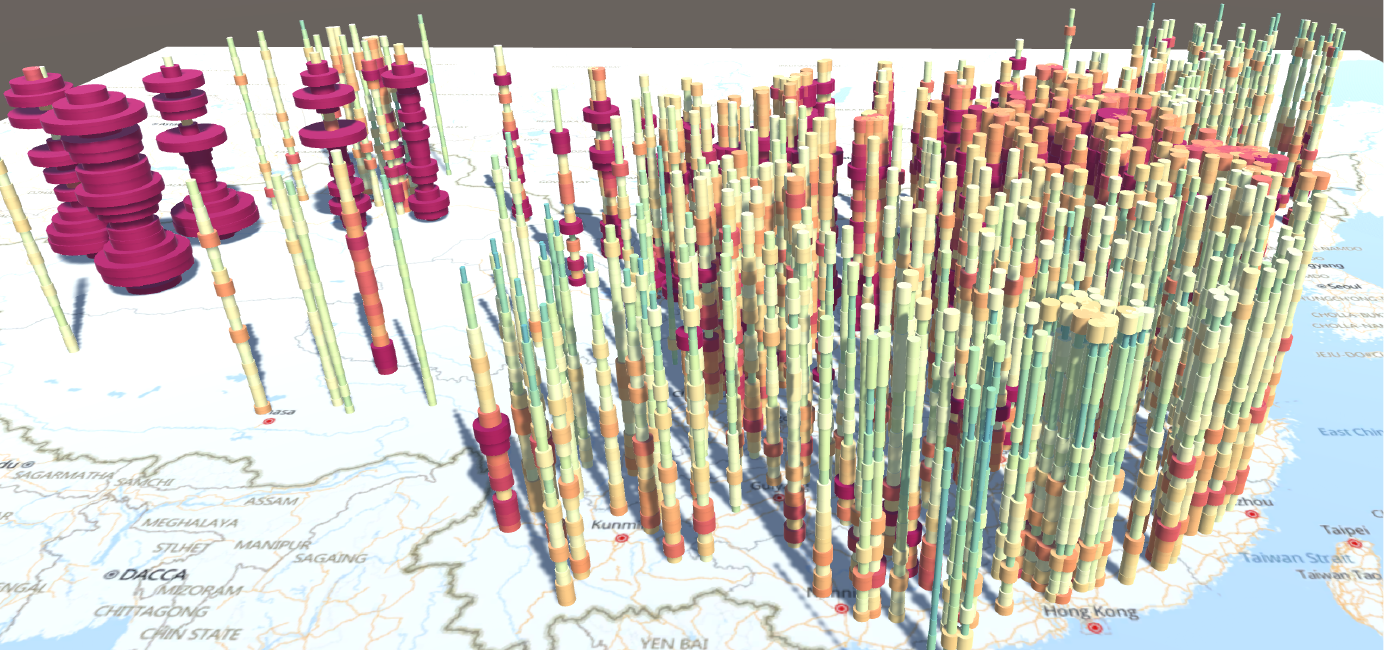}
  \caption{%
  	Our implementation of Thakur and Hanson's method~\cite{DBLP:conf/iv/ThakurH10}, later used as a baseline in the comparative user study.%
  }
  \label{fig:baseline}
\end{figure}

\subsection{Research Problem and Solution}
With the development of sensor technology and the decrease in storage costs, more and more fine-grained large-scale ST series are collected.
To analyze large-scale data adequately, analysts need effective and interactive visualization techniques.
However, as concluded in \autoref{sec:relatedwork}, although many ST series visualizations are designed to present temporal trends with the spatial context, they require users to make costly switches between geographic and temporal information, which is particularly unfriendly in the face of large-scale datasets.
The space-time cube is the most promising technique for alleviating the context switching cost.
Yet, existing attempts focused on small-scale datasets and may experience occlusion problems (\autoref{fig:baseline}).
Our research problem is refined to address how to adapt the space-time cube to large-scale ST series.

We realize that many spatiotemporal phenomena are essentially continuous in both the space and time domains, such as meteorological phenomena like rainfall, humidity, and temperature, and air pollution.
Due to the limited number of sensors and their limited monitoring capabilities, the recorded ST series of these phenomena exhibit discreteness in the time and space domains.
Targeting such a kind of ST series, we propose VolumeSTCube towards better visualization of large-scale ST series.
It first transforms the discrete ST series data into continuous space-time volumetric data following the strategy of the space-time cube.
Afterward, the volume-based visualization is designed and then equipped with a set of flexible interactions.
The combination of the space-time cube and 3D volume visualization effectively presents large-scale ST series with less context switching cost and visual occlusion.

\section{VolumeSTCube}
This section introduces VolumeSTCube, which visualizes large-scale ST series in a space-time cube, targeting expert analysts working with spatiotemporal data, such as climate change researchers and air pollution control specialists.

\begin{figure*}[tb]
  \centering
  \includegraphics[width=\linewidth]{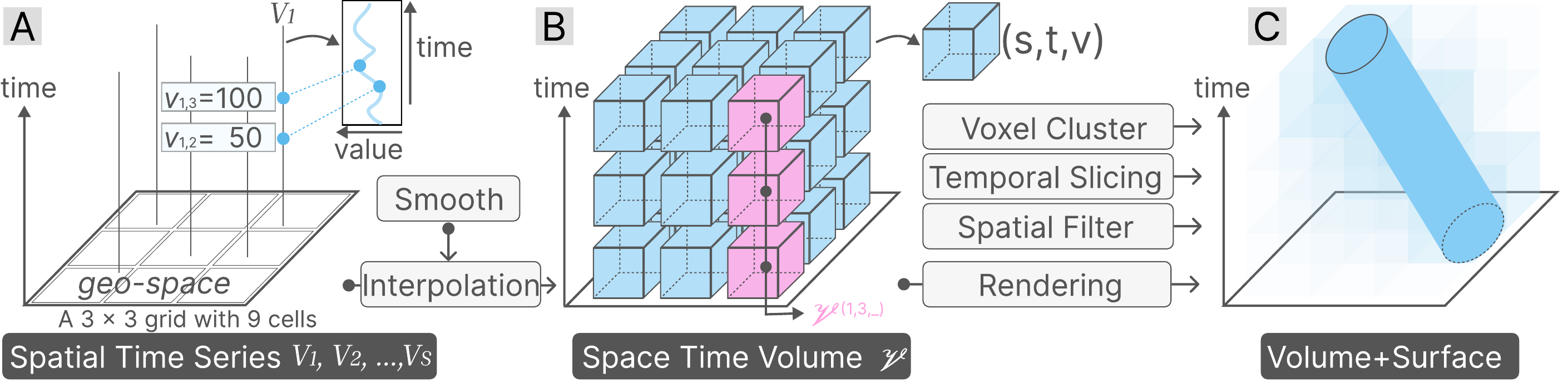}
  \caption{%
  	Pipeline of VolumeSTCube. (A) The spatial time (ST) series are placed in a 3D space following the space-time cube. (B) After interpolation and smoothing, the ST series are transformed as a space-time volume. (C) The volume is visualized with volume rendering and surface rendering, and is equipped with temporal slicing, spatial filter, and voxel cluster-based interactions, forming VolumeSTCube.%
  }
  \label{fig:pipeline}
\end{figure*}

\subsection{Pipeline}
The pipeline of VolumeSTCube is illustrated in \autoref{fig:pipeline}.

First, we generate a space-time cube space where the x-y plane represents the geographic space, and the z-axis represents the timeline.
Every ST series is represented as a column (i.e., the vertical line in \autoref{fig:pipeline}A) in the cube according to the geographic position.
The time-varying values of the ST series can be mapped to the column's vertical position.
Second, as shown in \autoref{fig:pipeline}B, the cube space is evenly divided into voxels, i.e., small regular cubes.
We apply interpolation and temporal smoothing to estimate the value $v$ for every voxel with a 3D position $(x, y, z)$.
As a result, all voxels together constitute a space-time volume where the values of adjacent voxels tend to be continuous.
Finally, the space-time volume is visualized with volume and surface rendering techniques (\autoref{fig:pipeline}C).
We also adapt the traditional 2D interactions into the 3D space-time cube environment, enabling users to effectively explore the volume with large-scale ST series.

VolumeSTCube is carefully designed based on the elementary space-time cube operations, outlined by Bach et al.~\cite{Bach2017STCInteraction}, including extraction, flattening, geometry transformation, and content transformation.
Specifically, the interpolation aligns with the filling of the content transformation operation.
Our visualization integrates filtering and shading as part of the content transformation operation and employs surface cutting as part of the extraction operation.
The user interactions are inspired by the volume-chopping of the extraction operation.

\subsection{Transformation}

We first leverage interpolation techniques to transform the ST series into a space-time volume. Further, smoothness is applied to the volume to enhance visual representation in the space-time cube.

\subsubsection{Interpolation}
First, we divide the geographic space into an $n \times m$ grid, resulting in $n \times m$ cells denoted as $C = \{c_{1,1}, c_{1,2}, \dots, c_{n,m}\}$.
Each cell is of uniform size and sufficiently small.
For example, the geographic space in \autoref{fig:pipeline}A is divided into a $3 \times 3$ grid with 9 cells.
Second, we divide the ST series with $T$ timestamps by their timestamps into $T$ slices.
For the $t$-th slice of the $t$ timestamp, $t \leq T$, there are observed data samples $Z_t = \{z_{1,t}, z_{2,t}, \dots\ z_{S,t} \}$, with $z_{i,t} = v_{i,t}$ representing the value of the $i$-th series at timestamp $t$.
Afterwards, interpolation techniques, like Kriging and Inverse Distance Weighted, can be employed to predict $\hat{z}_{c_{x,y}}$ for each grid $c_{x,y} \in C$ based on observed data samples $Z_t$, generating a sample $(s = (x,y), t, v = \hat{z}_{c_{x,y}})$ in the space-time volume.
For each slice of each timestamp, we obtain a 2D scalar field with $n \times m$ cells.
After performing the above operations for every slice, we stack the 2D scalar fields to form a completed space-time volume.

\subsubsection{Smoothing}
Each generated 2D scalar field is smooth over the geographic space because of the spatial autocorrelation principle.
However, due to many reasons, e.g., poor sensor quality, unstable data transmission, and the inherent uncertainty of interpolation, the space-time volume will inevitably have noises that manifest as sudden discontinuities in two 2D scalar fields at any adjacent timestamps.
The visual clutters caused by sudden discontinuities prevent users from performing analyses.

We employ a sliding average approach to reduce the temporal noise, inspired by the previous work~\cite{NBERmaca31-1}.
First, we partition the space-time volume $\mathcal{V}$ with a size of $n \times m \times T$ into $n \times m$ time series by the grid cell.
Each time series $\mathcal{V}^{x,y,\_}$, $1 \le x \le m$, $1 \le y \le n$, is an ST series at the cell $c_{x,y}$ with $T$ timestamps.
Subsequently, a sliding average smoothing is applied to every time series $V_{ij}$, respectively, given a window size.
The necessity of smoothing is illustrated in Appendix A.

\subsection{Visualization}
We employ volume rendering and surface rendering to visualize the space-time volume transformed from the ST series, aiming to reveal temporal trends, variation, and dynamics within the geographic context.
The contextual information regarding the spatial context and timeline is also provided in the cube space.

\textbf{Volume Visualization.}
We adopt raymarching~\cite{DBLP:journals/vc/Hart96} for volume visualization.
Firstly, a ray is cast from each pixel on the screen into the volume, incrementally advancing and sampling points along the ray.
Each sampled point carries color and opacity information based on the values of the voxels nearby, and the final color and opacity are computed using the Phong lighting model~\cite{10.1145/360825.360839}.
The contributions of sampled points along each ray are accumulated in a front-to-back manner to compose the final color and opacity of each screen pixel.
The compositing formulation is as follows:
$$\alpha_{a} = (1 - \alpha_{a}) \alpha_{s} + \alpha_{a}, \  C_{a} = (1 - \alpha_{a}) C_{s} + C_{a}.$$
$C_{a}$ and $\alpha_{a}$ denote the accumulated color and opacity, respectively, while $C_{s}$ and $\alpha_{s}$ denote the sample point's color and opacity, respectively.
Please refer to Ray et al.'s paper~\cite{DBLP:journals/tvcg/RayPSC99} for more details.
As a result, the larger the value, the redder and more opaque the color is, and conversely, the greener and more transparent the color.

Additionally, we introduce a voxel threshold $\lambda_v$.
Voxels whose values $< \lambda_v$ will be skipped during compositing.
Users can interactively adjust $\lambda_v$ via a slider.

\begin{figure}[tb]
  \centering
  \includegraphics[width=\columnwidth]{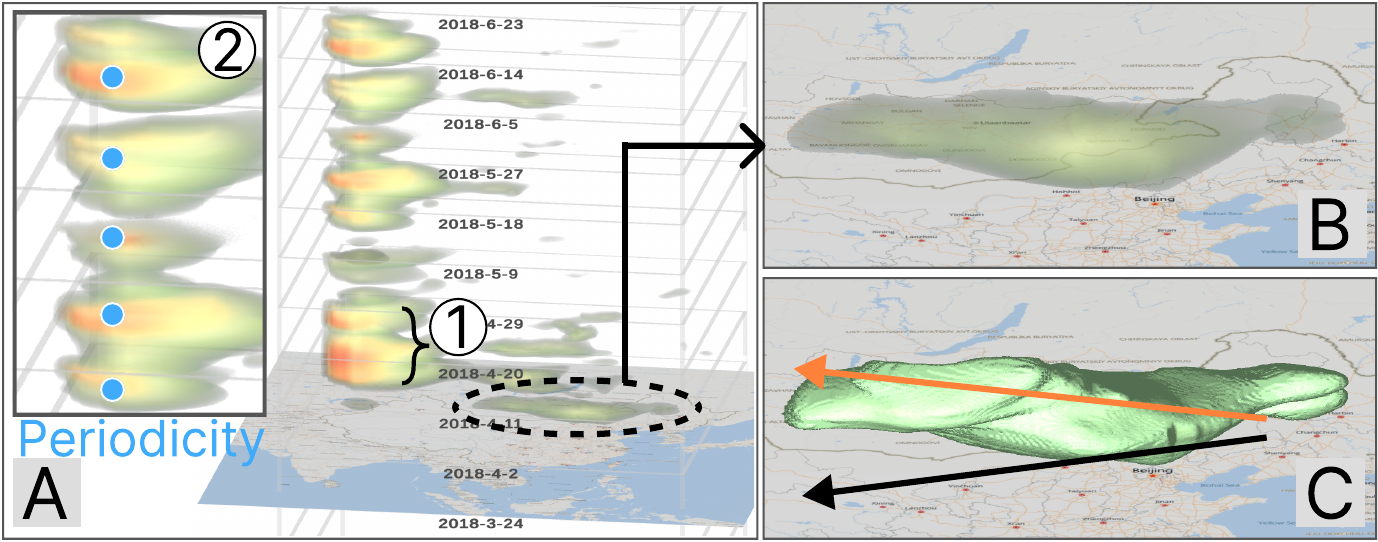}
  \caption{%
  	Visual designs of VolumeSTCube. (A) Space-time volume visualization with volume rendering and contextual information. (B and C) A propagation process of air pollution is visualized with volume rendering and surface rendering, respectively.%
  }
  \label{fig:vis}
\end{figure}

In this way, inherent spatiotemporal patterns in the ST series, such as temporal trends, spatiotemporal propagation~\cite{GeoNetverse}, and spatiotemporal evolution~\cite{DBLP:journals/jvis/YueFSLQH24}, can be easily analyzed based on the cluster of voxels with large values.
Assuming we are visualizing the ST series of air pollution concentration, below are some examples:
If a voxel cluster extends significantly in the vertical direction and over the x-y plane, it indicates that the duration of poor air quality persists across a vast spatial area (\autoref{fig:vis}A\textcircled{1}).
If in a vertical cylindrical area in 3D cube space are voxel clusters with equal spacing, the air pollution exhibits periodicity at that location (\autoref{fig:vis}A\textcircled{2}).
If the voxel cluster exhibits shifts in 3D cube space, it indicates a process of air pollution propagation (\autoref{fig:vis}B and C).

\textbf{Surface Visualization.}
The transparency of volume rendering will make it difficult to perceive spatiotemporal details.
To this end, we allow rendering isosurfaces given an iso-threshold $\lambda_i$ within the volume visualization above. By default, $\lambda_i$ = $\lambda_v$.
To do that, we only need to make the transparency of voxels with a value equal to $\lambda_i$ equal to 1.
That is, each voxel on the rendered surface has a value equal to $\lambda_i$.
Users can trigger the surface visualization and interactively adjust $\lambda_i$ via a slider. 
The surface in \autoref{fig:vis}C provides finer-grained details of the spatiotemporal pattern compared to \autoref{fig:vis}B.

\textbf{Contextual Information.}
We provide contextual information in VolumeSTCube, as shown in \autoref{fig:vis}A.
First, VolumeSTCube includes a geographic map perpendicular to the z-axis to enable the perception of the spatial context.
In addition, VolumeSTCube employs a box-shape z-axis. Each box indicates a time span, providing the time reference for the visual features in the vast space of the x-y plane.
The space-time cube space is evenly divided into multiple boxes along the z-axis.
The timestamp represented by the bottom of each box is displayed, with the text facing the user regardless of perspective rotation.

\subsection{Interactions}
In addition to the basic geometry transformation like rotation and translation, more flexible and user-friendly interactions are necessary for exploring a large-scale dataset.

\begin{figure}[tb]
  \centering
  \includegraphics[width=\columnwidth]{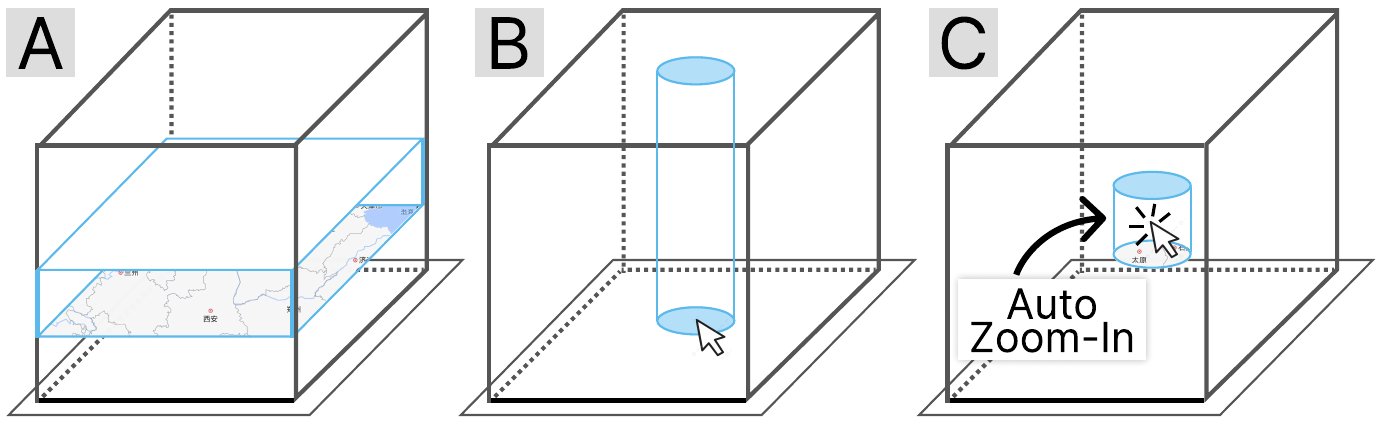}
  \caption{%
  	Interactions in VolumeSTCube. (A) Volume Slicing. (B) Volume Spotlight. (C) Voxel cluster-based Selection. %
  }
  \label{fig:interactions}
\end{figure}

\subsubsection{Considerations}
We reviewed linked views and integrated views in the 2D display and derived three kinds of interactions for spatiotemporal visualization.
Notably, the efficient ones focus on direct manipulation of spatial or temporal visualizations, instead of depending on context-free UI elements such as input fields and sliders.
Thus, we considered the tight integration of visualization and interactions in the interaction design.

First, users usually drill down into a time range of interest for analyzing a dataset with long-term observations.
They can specify the time range based on the focus+context mechanism~\cite{DBLP:journals/tits/WuWDBXWZDC21,10379518,DBLP:journals/tvcg/WangLZWHZXKZC23} or directly based on prior knowledge~\cite{DBLP:journals/tvcg/FerreiraPVFS13}.
Brushing on the visualization along the timeline aligns with user interaction practices in conventional 2D views.
A useful method is to integrate another 2D view in the 3D space for brushing~\cite{DBLP:conf/vr/FilhoSSN24}.
In contrast, we map the brushing interaction to the \textbf{volume slicing} along the timeline (\autoref{fig:interactions}A), allowing users to directly perform selection in the 3D space with less context switching.
Similarly, given a vast space where data is hard to analyze, users usually perform spatial range selection first~\cite{DBLP:journals/tvcg/LiuWL0ZQW17,DBLP:journals/tvcg/DengWLBZSXW22}.
Again, the selections with lassos and polygons on the map are difficult to issue in the 3D space.
Hence, we design an interaction called \textbf{volume spotlight} for users to select a spatial range in the cube space (\autoref{fig:interactions}B).

Besides, the time and spatial ranges can be selected simultaneously in recent visualization studies~\cite{deng2023visualizing}.
VolumeSTCube also enables the \textbf{spatiotemporal selection} (\autoref{fig:interactions}C).
With the aim of the patterns visualized in the space-time cube, users can directly and easily locate and select the time and spatial ranges.

\subsubsection{Volume Slicing for Time Range Selection}
VolumeSTCube enables users to select a time range by slicing the volume along the timeline, i.e., the z-axis.
To define this time range, the user first drags the geographic map, represented as an x-y plane, to a specific height along the z-axis corresponding to the starting point of the time range.
This action hides any voxels below the selected position.
Next, the user activates a second x-y plane and moves it to the endpoint of the time range, filtering out any voxels above this position.
This dual-plane approach slices the volume along the desired time interval, as illustrated in \autoref{fig:usecase}E1, F1, and G1.
While dragging each plane, the associated timestamp is displayed.
It also becomes easier for the user to relate the visualization to the spatial context as the map is closer to the visualization.

VolumeSTCube also supports users to drag two planes at once within a fixed time range via a time range slider.
In this way, it is easy for users to browse changes in spatial distribution over time.

\subsubsection{Volume Spotlight for Spatial Range Selection}
VolumeSTCube allows users to ``select'' a circular spatial range by placing a spotlight above the range on the cube space.
The specified circular range is defined by a center point and a radius. Based on these parameters, we can identify which voxels fall outside this area.
By adjusting the opacity of these outer voxels to zero, the focus remains exclusively on the voxels within the designated range, creating a spotlight effect, as demonstrated in \autoref{fig:usecase}C.

VolumeSTCube provides a user-friendly interaction that allows users to specify the circular range.
First, the user can directly specify the center of the range on the geographic map.
The map is an x-y plane perpendicular to the z-axis.
Given the mouse pointer's position on the screen, we can compute the intersection of the map plane and the ray emitted from the mouse position on the screen.
The center of the range is exactly the position of the intersection.
Second, the radius can be specified by scrolling the mouse wheel.
The range will be displayed as a black ellipse on the map for a visual hint.

\subsubsection{Voxel cluster-Based Spatiotemporal Selection}
The volume and surface are rendered as a whole.
Even if there are spatiotemporal patterns, it is difficult for users to select them directly.
We design a voxel cluster-based mechanism to enable the spatiotemporal selection.

\textbf{Voxel Cluster Detection.}
Analysts usually focus on sufficiently high values, such as abnormal hotspots or obvious propagation processes.
Thus, a voxel cluster in our study is defined as a spatiotemporal partition where the values of the voxels are larger than a threshold $\lambda_a$.
To detect the partitions, we first filter out the voxels whose values $\le \lambda_a$.
Afterward, we apply DBSCAN to cluster the remaining voxels based on their 3D positions.
DBSCAN is a density-based clustering algorithm suitable for partition detection because the voxels in each partition are close to each other.
Finally, each partition is considered a voxel cluster.

\textbf{Voxel Cluster-Based Interactions.}
For each remaining voxel, we additionally add a transparent but clickable hexahedron in the cube space according to the voxel's 3D position.
After the user clicks the hexahedron, the cluster the hexahedron belongs to is retrieved.
Consequently, the volume slicing and spotlight can be automatically executed to select the time and spatial ranges, respectively.
The time range is determined by the range of vertical positions over all voxels in the cluster.
The spatial range is a circle on the x-y plane that covers the projection of the cluster on the plane.

Users tend to click on the center of the cluster.
Therefore, the threshold $\lambda_a$ can be set slightly larger than expected in practice so that the detected clusters will have a smaller size, which prevents users from accidentally selecting by mistake.

\subsection{Implementation~\label{Sec:implementation}}
We implemented the data transformation and voxel cluster detection using Python 3.9. For interpolation, we employed Kriging, a widely used geostatistical method~\cite{DBLP:journals/tvcg/PrestonM23,Wackernagel1995}, leveraging the \textit{Pykrige} package. The DBSCAN algorithm was from the \textit{Scikit-learn} library.

VolumeSTCube is a desktop application. We choose Unity as our development platform over \textit{Three.js} based on WebGL. This decision considers that running in a browser may not meet the memory resource needs for volume rendering due to different browsers' memory management restrictions. The development environment is a desktop running Windows 10 with an Intel Core i7-13700K 3.40GHz CPU, NVIDIA GeForce RTX 3070 8GB GPU, and 32 GB of RAM.

\section{Evaluation}
VolumeSTCube is evaluated as follows.
First, we invite a professional analyst (PA) with five years of experience in analyzing nationwide air quality data to perform a real-world case study using VolumeSTCube.
Second, we conduct a controlled user study that compares VolumeSTCube with Thakur and Hanson's method~\cite{DBLP:conf/iv/ThakurH10} to further understand its advantages and disadvantages.

\begin{figure}[tb]
  \centering
  \includegraphics[width=\columnwidth]{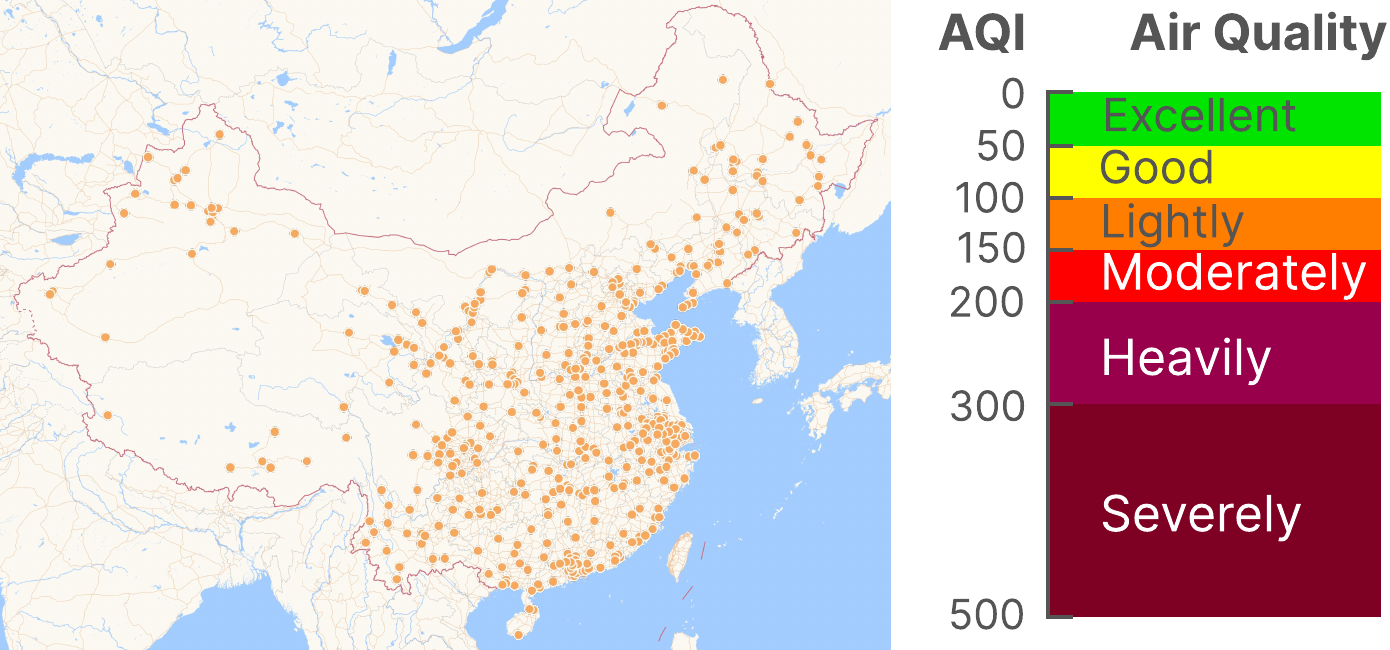}
  \caption{%
  (Left) Distribution of air quality stations. (Right) AQI and corresponding air quality descriptions.%
  }
  \label{fig:InterpolationSubRegions}
\end{figure}

\begin{figure*}[tb]
  \centering
  \includegraphics[width=\linewidth]{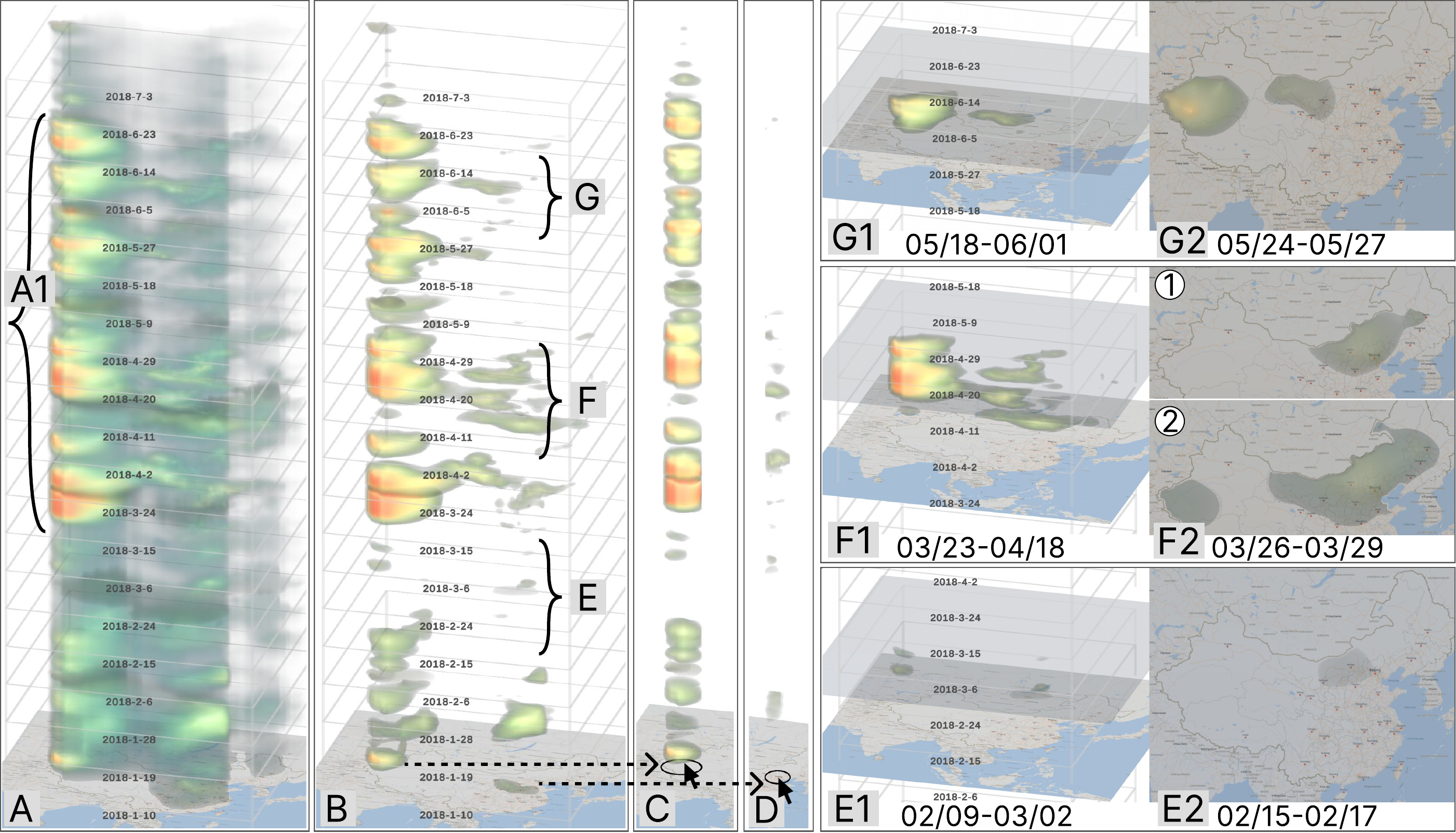}
  \caption{%
  	Case study of air quality in China. (A) VolumeSTCube provides the overview of air quality with $\lambda$ = 0. (B) The visual pattern becomes clearer with a larger $\lambda$ = 150. (C) The temporal variation of air quality in Xinjiang and (D) in the BTH region were revealed via volume highlight. (E, E1, and E2) The air pollution during the Spring Festival was generally below the moderate level. (F, F1, and F2) The air pollution caused by the dust storm from late March to April impacted a wide spatial range. (G) The air pollution caused by the dust storm in late May impacted a smaller spatial range.%
  }
  \label{fig:usecase}
\end{figure*}

We have a real-world AQI (air quality index) ST series dataset, collected from 448 air quality monitoring stations in China (\autoref{fig:InterpolationSubRegions}(left)).
In the whole evaluation section, the geographic space is divided into a 350 $\times$ 350 grid.
The time span is from January 1 to December 20, 2018, and the temporal granularity is one hour.
In sum, the dataset comprises 448 (stations) × 8,472 (timestamps) values, ranging from 0 to 500.
Each value in the dataset is the AQI at the timestamp of a monitoring station.
The higher the value, the worse the air quality.
In the data transformation procedure, the parameters of the Gaussian model are automatically determined by the \textit{PyKrige} library.
We evaluated the accuracy of the adopted Kriging interpolation on the dataset and found it to be acceptable, ensuring the reliability of subsequent analyses. For detailed results, please refer to Appendix D.
The parameters of DBSCAN, $\epsilon$ and $MinPts$, are configured as 10 and 100, respectively, after multiple trials.
The window size for smoothing is set as 24 hours.

The dataset is divided equally into two parts according to time, and each part is data for half a year.
The first dataset will be used in the case study, and both datasets will be used in the controlled user study.
Each of the datasets comprises 1.8 million values.

\subsection{Case Study: Air Quality in China}
We first introduced the visual encodings and interactions of VolumeSTCube.
Afterward, PA analyzed China’s air quality in the first half of 2018 via VolumeSTCube in person.
PA started with an overview and performed temporal, spatial, and spatiotemporal analyses.

\subsubsection{Overview}
After loading the dataset, PA obtained \autoref{fig:usecase}A, where the threshold $\lambda$ was 0.
The spatial and temporal variation of air quality could be roughly seen.
Some regions were particularly polluted during certain months, for example, the spatiotemporal partitions denoted in \autoref{fig:usecase}A1.
To make the representation clearer, PA increased the threshold $\lambda$ to 150.
AQI $>$ 150 means the air is moderately polluted, according to China's Ministry of Environmental Protection.
In this way, PA could also pay more attention to the occurrences of moderate and severe air pollution.

The result of increasing $\lambda$ was shown as \autoref{fig:usecase}B, which revealed some interesting macro patterns.
For instance, during the period of \autoref{fig:usecase}F, almost all of China experienced significant pollution.
In contrast, air pollution was lower during the period of \autoref{fig:usecase}E.
Besides, the frequent severe pollution in the west of China (shown in \autoref{fig:usecase}A1) became obvious with extensive yellow and red voxels.

\subsubsection{Temporal Analysis}
PA analyzed the temporal variation of AQI in western China, where PA observed frequent severe pollution, and
Beijing-Tianjin-Hebei (BTH) region, the capital economic circle of China.

\textbf{Western China.}
To analyze the air pollution situation in western China further, PA applied the volume highlight tool to select the spatial range of interest.
Recall that VolumeSTCube allows us to easily locate the spotlight on the target region by directly pointing to the region on the map.
The region was exactly the Xinjiang.
PA clearly observed that starting around March, multiple significant air pollution events occurred (\autoref{fig:usecase}C).
During these events, the air quality in Xinjiang deteriorated notably, possibly due to the poor atmospheric dispersion conditions and the influence of dust storms~\cite{LUO2020140560}.

\textbf{BTH.}
The BTH region was selected using the volume highlight tool (\autoref{fig:usecase}D).
Overall, there were only a few episodes of moderate or above pollution in the BTH region, and they did not last very long.
Moreover, starting from April, the overall air pollution in the BTH region consistently remained below the moderate level, i.e., AQI $\le 150$.

\subsubsection{Spatial Analysis} 
PA explored the air quality in China at some specific timestamps or time ranges.
First, PA was interested in the air pollution situation during the Spring Festival (\autoref{fig:usecase}E).
Besides, from late March to April (\autoref{fig:usecase}F), and in late May (\autoref{fig:usecase}G), two dust storms that were well-known to the whole country occurred in northern China.

\textbf{Spring Festival.}
The Spring Festival is during the time range of
\autoref{fig:usecase}E, where PA identified a period characterized by low air pollution levels nationwide, lasting more than half a month.
With volume slicing, PA selected the time range (\autoref{fig:usecase}E1).
Only a few voxels were observed.
This indicates that during this period, there was little moderate pollution across the country.
PA further narrowed the time range to New Year's Eve and the first two days of the new year.
Looking from top to bottom, PA saw that voxels only existed in north-central China and tended to be transparent (\autoref{fig:usecase}E2).
In other words, only north-central China had moderate air pollution during these three days.

The aforementioned observations were consistent with findings reported by the Ministry of Ecology and Environment of China: the air quality was generally good during the Spring Festival.
On New Year's Eve, as families gathered at home, the roads were devoid of traffic, resulting in reduced pollution emissions.
During this time, air pollution may be primarily attributed to fireworks.

\textbf{From Late March to April.}
During this period (\autoref{fig:usecase}F), multiple dust storms occurred and were reported by various media in China.
PA selected this time range by slicing the volume again (\autoref{fig:usecase}F1).
Severe pollution occurred in Xinjiang as voxels were very red.
Moreover, moderate pollution was present in large areas besides Xinjiang.
PA wanted to analyze how large an area was affected by this dust storm.
To do that, PA adjusted the map and the upper plane and specified a three-day (from March 26 to March 29) time range during which air pollution occurred in vast areas.
\autoref{fig:usecase}F2\textcircled{1} was the top view for this time range.
Due to the dust storm, a large region of northern China suffered from moderate to above-average pollution.
The region included Beijing, the capital of China, and three province capitals, Hoerhot, Harbin, and Taiyuan.

Furthermore, PA reduced $\lambda_v$ to 100 to obtain the areas that were lightly polluted due to the dust storm (\autoref{fig:usecase}F2\textcircled{2}).
The impact of this storm covered the three northeastern provinces, the Loess Plateau, and half of the North China Plain.

\textbf{Late May.}
\autoref{fig:usecase}G1 showed the air pollution caused by this dust storm.
PA also specified a one-day time range and analyzed the spatial impact of this storm.
\autoref{fig:usecase}G2 was the top view of the visualization during this time range.
Mainly, the regions around the Taklimakan Desert and the Badain Jaran Desert were moderately polluted due to the dust storm.
The spatial range of moderate pollution caused by this dust storm is smaller than that caused by the abovementioned dust storm.

\begin{figure}[tb]
  \centering
  \includegraphics[width=\columnwidth]{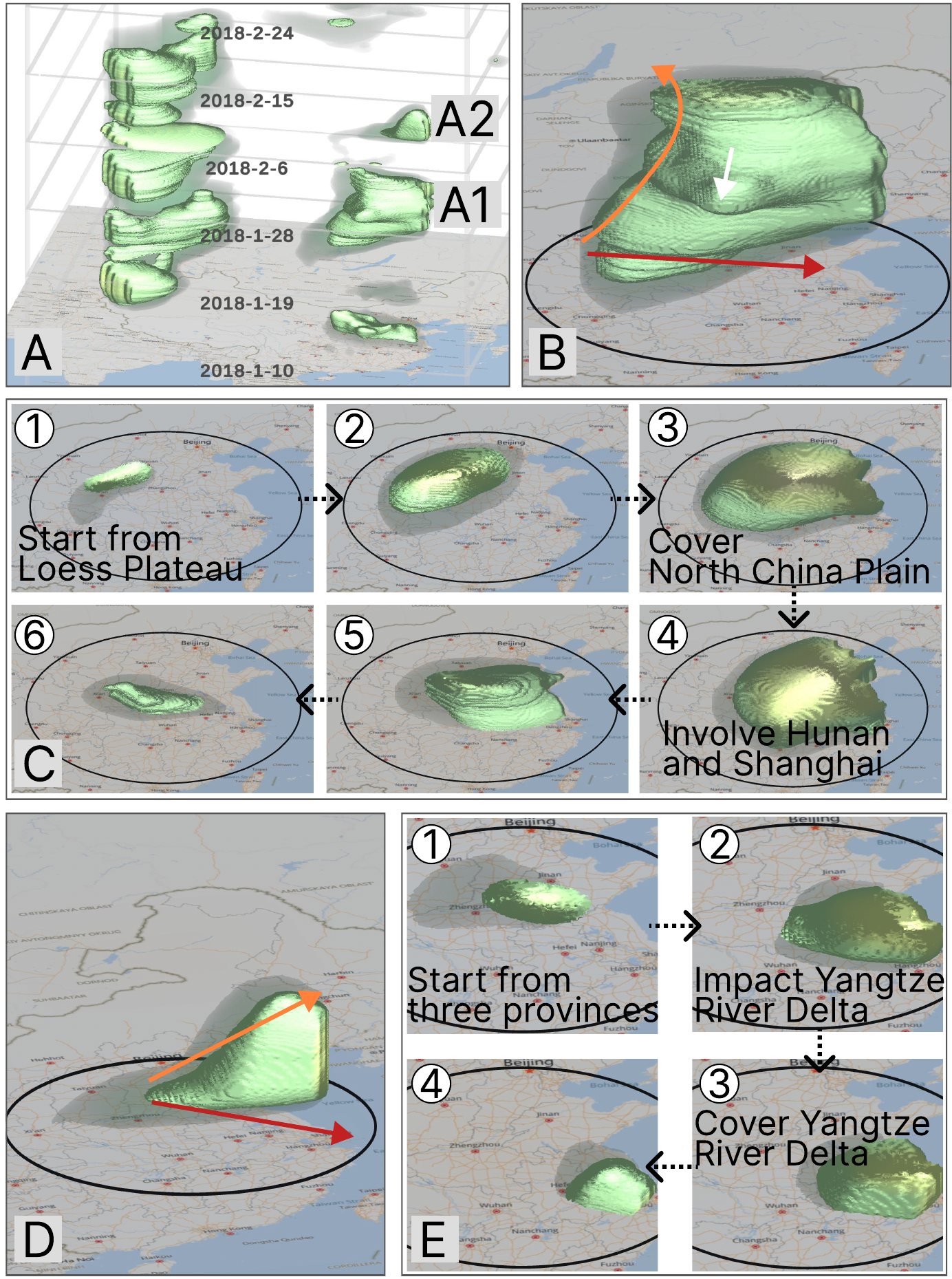}
  \caption{%
  	Case study of air quality in China. (A) Air quality data is visualized with the surface rendering in VolumeSTCube, where (A1 and A2) two propagation processes of air pollution can be identified. (B) After clicking (A1), the user can obtain this view. (C) The air pollution propagated from the Loess Plateau to the North China Plain. (D) After clicking (A2), the user can obtain this view. (E) The air pollution propagated from the North China Plain to the Yangtze River Delta.%
  }
  \label{fig:usecase2}
\end{figure}

\subsubsection{Spatiotemporal Analysis}
VolumeSTCube can also support spatiotemporal analysis well because it enables perceiving time and space simultaneously.
The evidence was that PA could quickly identify many \textbf{hotspots} in \autoref{fig:usecase}B and C.
Beyond hotspots, PA could also easily identify the \textbf{propagation process} of air pollution, which expands the impact of air pollution over geographic space and time.

\textbf{Propagation from Loess Plateau to North China Plain.}
In the bottom portion of \autoref{fig:usecase}A, a voxel cluster was evident, exhibiting a tilted 
configuration within 3D space-time cube space.
Such a tilted shape signified the spatiotemporal propagation process of air pollution.
With rendered isosurfaces ($\lambda_i$=150 and $\lambda_v$=125) (\autoref{fig:usecase2}A), PA could clearly discern the propagation process.
To analyze the process in detail, PA clicked the surface (\autoref{fig:usecase2}A1), i.e., the voxel cluster detected with $\lambda_a = 150$.
As a result, spatial selection, temporal selection, and zooming-in functionalities were automatically issued, thereby eliminating the clutter from non-selected visual elements.
From \autoref{fig:usecase2}B, PA obtained the following three pieces of information.
First, the pollution generally propagated toward the southeast, as indicated by the red arrow.
Second, the propagation process appeared to be a round-trip, as indicated by the orange arrow.
Third, the pollution once spread to the southwest, as indicated by the white arrow.
Identifying these visual patterns was attributed to the surface rendered with lighting information.

PA narrowed down the time range to one day and moved the range along the timeline to investigate the fine-grained propagation process.
Some snapshots were presented in \autoref{fig:usecase2}C\textcircled{1}-\textcircled{6}.
Particularly, the moderate pollution started from the Loess Plateau (\autoref{fig:usecase2}C\textcircled{1}).
Then, the North China Plain (\autoref{fig:usecase2}C\textcircled{3}), Hunan, and Shanghai (\autoref{fig:usecase2}C\textcircled{4}) were successively affected by pollution.
Air pollution in Hunan and Shanghai disappeared early (\autoref{fig:usecase2}C\textcircled{5} and \textcircled{6}).

\textbf{Propagation from North China Plain to Yangtze River Delta.}
From \autoref{fig:usecase2}A, PA also noticed another propagation process, enclosed in \autoref{fig:usecase2}A2.
After clicking it, PA obtained \autoref{fig:usecase2}D.
Pollutants spread to the southeast again, towards the red arrow.
In addition, the rendered object had a flat base, which meant the propagation process was rapid.
The air pollution first occurred at the junction of Henan, Shandong and Anhui provinces (\autoref{fig:usecase2}E\textcircled{1}), quickly affected (\autoref{fig:usecase2}E\textcircled{2}), and covered the whole Yangtze River Delta (\autoref{fig:usecase2}E\textcircled{3}), and eventually dissipated in the Yangtze River Delta (\autoref{fig:usecase2}E\textcircled{4}).

The aforementioned propagation processes occurred around January 20 and February 3, 2018, respectively.
Concurrently, China experienced two significant cold waves that resulted in strong cold air moving southward, leading to a substantial drop in temperatures.
PA explained these meteorological conditions likely contributed to the observed patterns of air pollution.

\subsubsection{Case Study Summary}
This case study demonstrates VolumeSTCube's effectiveness in supporting large-scale spatiotemporal analysis oriented to space, time, and even spatiotemporal patterns.
Particularly, the visualization based on the space-time cube enables users to directly map temporal trends to geographic space, facilitating space-oriented analysis.
Moreover, VolumeSTCube does not require the trial-and-error parameter adjustment process, allowing for the direct revelation of spatiotemporal patterns through spatial and temporal integrated presentations.

\begin{table*}[t]
\centering
\caption{%
Tasks designed for the user study, 
involving lookup, comparison, and relation-seeking at elementary and synoptic levels, covering space-oriented (\sethlcolor{lightBlue}\hl{light blue} background), time-oriented (\sethlcolor{lightRed}\hl{light red} background), and spatiotemporal pattern-oriented (\sethlcolor{lightGreen}\hl{light green} background) analyses.%
}
\begin{tabular}{l|ll}\label{tab:tasks}
 & Elementary Tasks & Synoptic Tasks \\ \hline
Lookup & \cellcolor[HTML]{bceaf7}\begin{tabular}[c]{@{}l@{}}Q1: At a given timestamp, where does (not) severe pollution occur?\end{tabular} & \cellcolor[HTML]{bceaf7}\begin{tabular}[c]{@{}l@{}}Q2: At a certain timestamp, how much area suffers from \\ severe air pollution?\end{tabular} \\ 
 & \cellcolor[HTML]{ffdede}Q3: In a given region, when dose pollution occur? & \cellcolor[HTML]{ffdede}\begin{tabular}[c]{@{}l@{}}Q4: In a given city, what are the pollution development\\ and dispersion trends?\end{tabular} \\
 & \cellcolor[HTML]{e3f6bc}Q5: How is the air pollution in a given city at a certain time? & \cellcolor[HTML]{e3f6bc}Q6: Where and when are hotspots? \\ \hline
Comparison & \cellcolor[HTML]{bceaf7}\begin{tabular}[c]{@{}l@{}}Q7: At a given timestamp, region A or region B, which one \\ has more severe air pollution?\end{tabular} & \cellcolor[HTML]{bceaf7}\begin{tabular}[c]{@{}l@{}}Q8: At a given timestamp, which region is \\ the most seriously polluted?\end{tabular} \\
 & \cellcolor[HTML]{ffdede}\begin{tabular}[c]{@{}l@{}}Q9: In a given region, during period A or period B, which period\\ has more severe air pollution?\end{tabular} & \cellcolor[HTML]{ffdede}\begin{tabular}[c]{@{}l@{}}Q10: In a given city, during which period did pollution\\ accumulate the fastest?\end{tabular} \\
 & \cellcolor[HTML]{e3f6bc}Q11: When and where did the most severe pollution occur? & \cellcolor[HTML]{e3f6bc}Q12: which hotspot is the largest? \\ \hline
Relation-seeking & \cellcolor[HTML]{bceaf7}\begin{tabular}[c]{@{}l@{}}Q13: At a given timestamp, which two cities suffer from\\ severe pollution simultaneously?\end{tabular} & \cellcolor[HTML]{bceaf7}\begin{tabular}[c]{@{}l@{}}Q14: At a given timestamp, which regions suffer from \\severe air pollution simultaneously?\end{tabular} \\
 & \cellcolor[HTML]{ffdede}\begin{tabular}[c]{@{}l@{}}Q15: In a given city, did any severe pollution happen within\\ one month before a given pollution?\end{tabular} & \cellcolor[HTML]{ffdede}\begin{tabular}[c]{@{}l@{}}Q16: In a specific city, is there a periodical pattern to\\ the occurrence of air pollution?\end{tabular} \\
 & \cellcolor[HTML]{e3f6bc}\begin{tabular}[c]{@{}l@{}}Q17: Are there two close cities experiencing severe pollution\\ one after another?\end{tabular} & \cellcolor[HTML]{e3f6bc}\begin{tabular}[c]{@{}l@{}}Q18: How is the propagation process? \\ (Describe the propagation over space and time)\end{tabular}
\end{tabular}
\end{table*}

\subsection{User Study}
We conducted a controlled, within-subject user study.
The study primarily aimed to verify 1) the effectiveness of continuous volume-based visualization compared to column-based visualization and 2) the ease of understanding of volume-based visualization combined with the space-time cube.
Besides, we hoped to identify the strengths and weaknesses of VolumeSTCube.
While numerous methods, such as 2D map-based animations or coordinated 2D maps and line charts, are available for comparison, these approaches differ fundamentally in design and interaction principles from STC-based methods. We selected Thakur and Hanson’s STC-based method~\cite{DBLP:conf/iv/ThakurH10} as the baseline, shown in \autoref{fig:baseline}. This method, the most recent STC-based visualization for ST series, lacks volumetric representation but serves as a representative of similar methods~\cite{Tominski2005STC,DwyerGallagher2004STC}. Its design allows a focused evaluation of the benefits introduced by volume-based visualization.

\subsubsection{Study Setup}

\textbf{Subjects.}
Due to the challenge of recruiting a sufficient number of domain experts, we opted to involve undergraduate students.
If undergraduate students, who are less familiar with advanced visualization techniques and domain-specific problems, were able to complete the tasks successfully after the introduction or tutorials, it stands to reason that domain experts, with their greater expertise and experience, would be even better equipped to utilize the system effectively. Finally, we invited twelve undergraduate students (six males and six females) as subjects.
Their majors include software engineering (4 subjects), business administration (2), product design (2), logistics management (1), industrial design (1), tourism management (1), and journalism (1).
They possess a common-sense understanding of air pollution but relatively limited visualization and spatiotemporal analysis expertise.
These subjects are suitable for the purposes of our study.

\textbf{Tasks.}
Andrienko and Andrienko~\cite{DBLP:books/daglib/0015278} classified general and basic spatiotemporal analytical tasks into lookup, comparison, and relation-seeking at elementary and synoptic levels.
We follow this well-established taxonomy to design concrete tasks, as Thakur and Hanson~\cite{DBLP:conf/iv/ThakurH10} did.
We designed 18 tasks, as shown in \autoref{tab:tasks} with the question form.
These tasks cover the three kinds of tasks (i.e., lookup, comparison, and relation-seeking) at two levels (i.e., elementary and synoptic) proposed in Andrienko and Andrienko's taxonomy~\cite{DBLP:books/daglib/0015278}.
Elementary tasks deal with the elements of data, while synoptic tasks are performed on the spatiotemporal pattern rather than the elements.
Considering our scenario and dataset, we define patterns as spatiotemporal hotspots and propagation processes.
Moreover, we further refine the tasks from the spatial, temporal, and spatiotemporal perspectives, as indicated by the blue, pink, and green backgrounds, respectively.
Since the subjects were not experts in air quality, the tasks only required them to describe the spatial, temporal, or spatiotemporal phenomena objectively and did not involve explaining the phenomena, such as how pollutants were generated.

\textbf{Baseline.}
In the baseline (\autoref{fig:baseline}), each column represents an ST series and is positioned on the map according to its geographic location.
Disks distributed along the z-axis within each column visualize the values at different timestamps.
The size and color of the disks encode the magnitude of the values.
Larger disks and redder colors indicate higher values and smaller disks and greener colors indicate lower values.

In Thakur and Hanson’s implementation using U.S. food stamp and unemployment rate data~\cite{DBLP:conf/iv/ThakurH10}, occlusion is minimal due to the lower frequency of timestamps resulting from aggregation, with only one spatiotemporal (ST) series per state.
Our dataset demands a careful balance in aggregation; broad time ranges may lead to the loss of critical details. Therefore, we set the aggregation interval to 24 hours, where each disk represents the average air quality for a single day. 
In addition, we preserved the original spatial distribution of the ST series to reflect the complexities of large-scale data scenarios accurately.

We implement the column selection interaction described in Thakur and Hanson's paper~\cite{DBLP:conf/iv/ThakurH10}, which corresponds to our volume spotlight for spatial selection. 
Additionally, we equip the baseline visualization with the time range selection interaction similar to VolumeSTCube, enabling a fair comparison between the two methods.

\textbf{Procedure.}
Firstly, we introduce the key concepts, including ST series, air quality datasets, common temporal trends, spatial distributions, and prevalent spatiotemporal patterns within air quality data. 
Subsequently, we provide the subjects with a comprehensive introduction to the system's visual encoding and interactions, followed by a hands-on exploration of a sample dataset using the system.
Next, we proceed with the formal experiment section.
Each subject analyzed the dataset and answered questions for the first and second halves of the year (denoted as Dataset A and Dataset B, respectively) using two different systems.
Upon completion of the experiment, we gather the subject feedback through a brief interview.
The procedure for each subject lasted about 1 hour, with a 5-minute break after using the first system.

Specifically, in the experiment section, the twelve subjects were randomly assigned into four groups based on the order of system usage and the corresponding dataset (e.g., VolumeSTCube with Dataset B, baseline with Dataset A).
18 questions per dataset were instantiated based on the dataset according to the 18 tasks in \autoref{tab:tasks}.
These questions were then reorganized according to the exploration process, grouping together those that could be answered sequentially to minimize the workload on the subjects.
The system would be reset between different question groups, irrespective of which system the subject was utilizing.
Please refer to Appendix C for the questions, their order and grouping, correct answers, and correctness criteria.
Note that regardless of which system you use, the order in which the questions appear is the same.
When the subjects were conducting the experiment, the questions were displayed on a tablet.
Once the subject confirmed understanding the question, the subject used the system on the desktop for analysis.
The response time was recorded from the moment of the confirmation to the moment the answer was provided.
During the experiment, a think-aloud protocol is employed, encouraging subjects to verbalize their thoughts.

\begin{figure*}
  \centering
  \includegraphics[width=\linewidth]{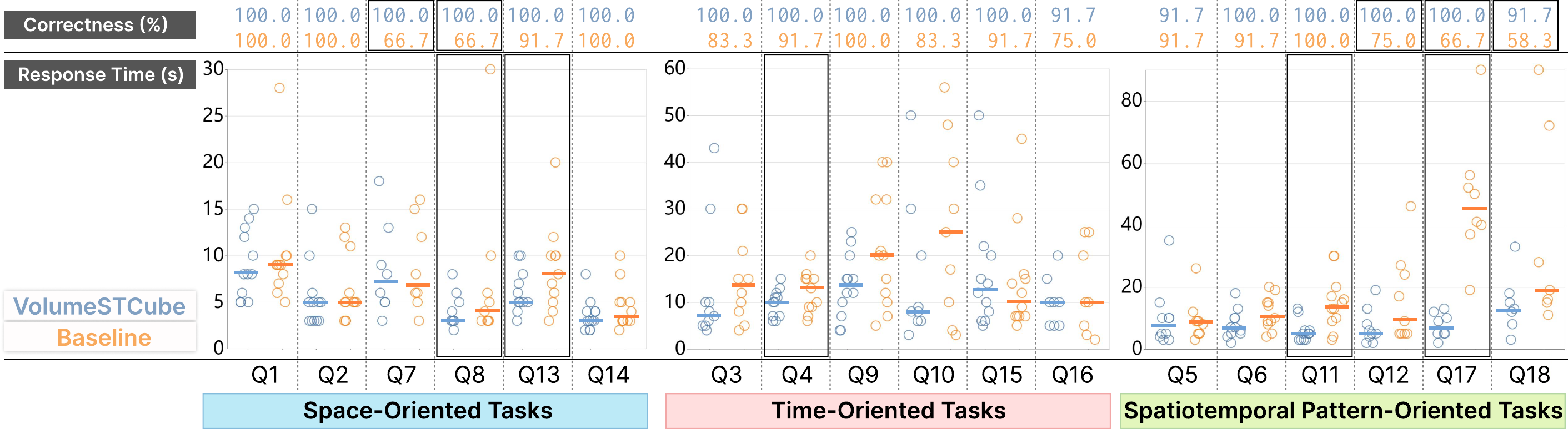}
  \caption{%
    Results of user study. 
    The results for VolumeSTCube and the baseline are represented by soft blue and soft orange, respectively. The bottom section categorizes the 18 questions into three groups based on their spatial, temporal, and spatiotemporal perspectives. The middle section displays the response times for each question with jittered dot plots, where horizontal markers denote medians. The top section shows the correctness of the answers. Black rectangles highlight instances where either the difference of the correctness is more than 25\% or the response time is significantly different. No instances were found where the baseline obviously outperforms VolumeSTCube.
    }
  \label{fig:UserStudy}
\end{figure*}

\subsubsection{Results}
If the subject chose to give up, it was also counted as an incorrect answer.
In addition, response records corresponding to incorrect answers, as well as the user's response records for the same question when using the alternate system, will be excluded from the response time calculations.
\autoref{fig:UserStudy} summarizes the correctness of the subjects' answers and the subjects' response times for every question.

Overall, \autoref{fig:UserStudy} suggests that VolumeSTCube outperforms the baseline, with advantages in eight questions, as highlighted by the black rectangles.
Regarding the answer correctness, using VolumeSTCube is at least 25\% more accurate than using the baseline on Q7, Q8, Q12, Q17, and Q18.
Regarding the response time, VolumeSTCube significantly outperforms the baseline in the support for Q4 ($p < 0.05$), Q8 ($p < 0.05$), Q11 ($p < 0.01$), Q13 ($p < 0.05$), and Q17 ($p < 0.01$) based on the Wilcoxon test's results.
No instances were observed where the baseline obviously outperforms VolumeSTCube.
Below, we elucidate the performance of both the baseline method and VolumeSTCube in supporting subjects to answer questions Q7, Q8, Q11, Q13, Q12, Q17, and Q18.

\textbf{Q7, Q8, and Q13.}
These three questions required the subjects to examine the value distribution across the geographic space at a given timestamp. 
To do that, the slicing interaction was helpful regardless of the system used.
Consequently, the response times were typically under 20 seconds, with most being under 10 seconds.
However, in the baseline, visualizations were influenced by station density, leading some subjects to mistakenly perceive areas with dense disk density as more polluted, resulting in an accuracy of only 66.7\% of Q7 and Q8.
VolumeSTCube employs a continuous visual representation (e.g., \autoref{fig:usecase}E2, F2, and G2) to mitigate this confusion.

\textbf{Q11.}
To answer Q11, subjects are required to explore all timestamps across the entire geographic area.
Using VolumeSTCube, subjects can examine all voxels without encountering occlusion issues and easily identify the reddest and most opaque ones through simple rotations (only a few seconds).
In contrast, the baseline demands subjects repeatedly select time ranges and columns to mitigate occlusion caused by columns, thereby prolonging interaction times.

\textbf{Q12.}
To solve Q12, subjects should identify the largest hotspot.
In VolumeSTCube, identifying hotspots is expedited as they are presented as distinct voxel clusters. 
Subjects can easily compare the size and color of each voxel cluster, and thereby identify the largest one.
Conversely, subjects using the baseline method found it challenging to compare hotspots due to occlusion, especially in densely clustered columns in North China, as shown in \autoref{fig:baseline}.
They resorted to selecting time ranges and columns to estimate the largest hotspot, which was time-consuming and imprecise.
Consequently, the baseline takes 8 seconds longer on average than VolumeSTCube.
Furthermore, with the baseline method, one subject gave up, and two subjects provided incorrect answers, resulting in an accuracy rate of 75\%, which is lower than VolumeSTCube's accuracy rate of 100\%.

\textbf{Q17.}
In the baseline, the propagation process is hidden in multiple discretely distributed columns and subjects had to move the time range back and forth and observe changes in the color or radius of every disk, which is time-consuming with the mental burden. 
Nonetheless, subjects using VolumeSTCube can easily identify a propagation process via the rendered surface in a continuous form (e.g., shown in \autoref{fig:usecase2}) and select two cities based on the propagation path.
As a result, VolumeSTCube outperformed the baseline by over 40 seconds regarding response time; two subjects gave up when using the baseline, resulting in a correctness of 66.7\%.

\textbf{Q18.}
As for Q18, subjects were tasked with describing a propagation process.
Similar to Q17, the continuous surfaces in VolumeSTCube help subjects to describe the propagation process (e.g., shown in \autoref{fig:usecase2}).
Subjects using the baseline took an average of 21 seconds longer than those using VolumeSTCube.
Notably, four subjects gave up, and the accuracy rate was only 58.3\%, considerably lower than VolumeSTCube's 91.7\%.

The above questions (tasks) are either space-oriented or spatiotemporal pattern-oriented, but through observation and test, VolumeSTCube has advantages on time-oriented tasks as well compared to the baseline.
For Q3, Q4, Q10, Q15, and Q16, the correctness of subjects' answers using VolumeSTCube was slightly higher compared to using the baseline method.
The average response time for questions Q9 and Q10 was at least 7 seconds shorter when using VolumeSTCube than when using the baseline.
For Q4 and Q9, the Wilcoxon test on the response times of using the two systems results in $p$ = 0.03 (significance) and $p$ = 0.06 (weak significance), respectively.
These observed results may be attributed to a preference for continuous visual representations along a timeline (e.g., \autoref{fig:usecase}C), which allows for clear and fine division of events, as opposed to discrete visual elements such as stacked disks.

\subsubsection{User Feedback~\label{Sec:feedback}}
All subjects appreciated VolumeSTCube, considering it highly easy-to-read and efficient for analyzing the air quality dataset.
Their feedback is summarized into visualization and interaction aspects:

\textbf{Visualization.}
All subjects quickly grasped the space-time cube and confirmed the intuitiveness of VolumeSTCube.
Moreover, all subjects using VolumeSTCube for analysis experienced minimal occlusion problems and remarked that ``\textit{due to the opacity of voxels (actually the volume visualization they meant), they experienced no visual occlusion during analysis}.''
Two subjects noted that the opacity of voxels could potentially hinder analyzing low pollution levels, which corresponds to VolumeSTCube's poor performance in Q15.
In the future, we plan to implement a rendering parameter adjustment control to alleviate this problem.
For example, the mapping of values to transparency and color can be adjusted interactively.
Finally, most subjects expressed that surface rendering offered clear boundaries for voxel clusters, aiding in obtaining specific values at spatiotemporal positions and assessing propagation processes.
This capability could be difficult to achieve solely with volume rendering.
Interestingly, some subjects with strong spatial perception felt relying solely on volume visualization was sufficient.

\textbf{Interaction.}
All subjects stated that the interactions regarding the spatial range and temporal range selections implemented in the two systems were helpful and effective in exploring large-scale spatiotemporal data.
More interestingly, most subjects took it for granted that the voxel clusters in VolumeSTCube were clickable or could be selected.
VolumeSTCube's voxel cluster-based interaction exactly offers such interactivity.
Several subjects exactly commented, ``\textit{I prefer the voxel cluster-based interaction to locate hotspots in one step rather than the volume slicing first, followed by the volume spotlight and zooming-in}.''

\section{Discussion}\label{sec:discussion}
This section discusses the implications, generalizability, and limitations of VolumeSTCube, and posts the future work.

\subsection{Implications}
The STC, a form of 3D visualization, has historically seen mixed reception in the visualization community, particularly due to inherent occlusion issues~\cite{DBLP:series/lncs/Munzner08}.
Nonetheless, our study shows that the STC's effectiveness can be enhanced by reducing occlusion through data transformation combined with volume and surface visualizations.
Furthermore, to improve the STC's adaptability to large-scale datasets, we design interactions for users to manipulate the STC from time and space dimensions, corresponding to spatial and temporal selection interactions common in 2D views.
In the case study, the expert comprehensively analyzed air quality in China, demonstrating VolumeSTCube's effectiveness. The user study employed carefully designed tasks that required participants to utilize both volume and surface visualizations, alongside interactions such as volume spotlight, slicing, and voxel cluster-based techniques, to complete the analyses. The high accuracy and efficiency observed in task completion indicate that participants mastered VolumeSTCube, further underscoring its usability, effectiveness, and broad applicability.

\subsection{Generalizability}
VolumeSTCube can be generalized to ST series representing natural phenomena in various domains, as they are usually continuous over space and time. Representative phenomena include rainfall, humidity, and temperature, as well as air pollution.
The three modules of VolumeSTCube—transformation, visualization, and interaction—are designed to be domain-independent without any domain-specific constraints.
As long as phenomena exhibit continuity across both time and space, we can generate volumetric data through interpolation and smoothing, so that our visualizations and interactions can be applied.
In this study, the air pollution ST series is used to demonstrate.
In Appendix B, we include another example, where we visualize the temperature ST series using VolumeSTCube and briefly describe the observed spatiotemporal patterns. For more details, please refer to Appendix B.

VolumeSTCube is not suitable for ST series with discrete spatial distributions, such as tourist visits to different attractions or demographic trends across countries. 
For example, a country located between two populous countries may have a significantly smaller population.
Due to their spatially discrete nature, converting these types of ST series into continuous volumetric data would be ineffective and misleading.
Topological analysis that extracts spatially discrete events can be an effective alternative for supporting visual exploration of these datasets~\cite{DBLP:journals/tvcg/DoraiswamyFDFS14}.

\subsection{Scalability}
The scalability of VolumeSTCube is primarily constrained by the hardware's capacity to handle a certain number of voxels.
Under the hardware conditions mentioned in~\autoref{Sec:implementation}, VolumeSTCube can effectively visualize ST series spanning half a year across China, with a spatial granularity of a 350 $\times$ 350 grid and a temporal granularity of hours, totaling over 50 million voxels.
In cases where hardware resources are limited, adjustments such as reducing the number of cells or aggregating the time can be implemented to mitigate the number of voxels while maintaining the representation of ST series within the same range.

The evaluation demonstrates the effectiveness of the interactive volume visualizations provided by VolumeSTCube for ST series containing millions of records. Particularly, the data transformation module enables VolumeSTCube support for even larger datasets. For example, in the context of air pollution analysis, if there are 3,000 monitoring stations across China, the number of hourly records over six months amounts to approximately 12 million. The module can still transform the records for each hour into a 350 $\times$ 350 grid again, allowing VolumeSTCube to visualize these 12 million records with interactive volume visualizations effectively.

\subsection{Limitations and Future Work}
In addition to the lack of a rendering parameter control mentioned in \autoref{Sec:feedback}, we also identify the following limitations or future work.

\textbf{Uncertainty Visualization.}
Given the inherent uncertainty in the interpolation process, uncertainty visualizations are desirable for reliable analysis, helping assess whether observed patterns are uncertain and if local but important patterns are being overlooked~\cite{DBLP:journals/tvcg/PrestonM23}.
In the future, we plan to incorporate uncertainty into VolumeSTCube, using methods such as the uncertain volume visualization technique.

\textbf{Virtual/Augmented Reality.}
VR (Virtual Reality) and AR (Augmented Reality) technologies enable users to perceive, interact with, and analyze data through embodied visualizations and interactions, thereby enhancing their perception, cognitive abilities, and engagement~\cite{DBLP:journals/vi/ZhangWZST23,DBLP:journals/tvcg/BoorboorCKCBPK24}.
Inspired by the benefits of VR/AR and the rapid development of VR/AR technology, we plan to extend VolumeSTCube into a VR/AR environment.

\textbf{Viewpoint Selection.}
In 3D space, a good viewpoint allows users to see patterns clearly~\cite{10478337}.
At present, the user needs to manually move and rotate the camera to find an ideal viewpoint.
In the future, we plan to leverage viewpoint selection methods to locate the camera.
Furthermore, camera movement optimization will be incorporated to enhance the navigation and narrative.

\textbf{Advanced Spatiotemporal Analysis.}
VolumeSTCube should be further extended to facilitate comparative analysis and the examination of complex spatiotemporal patterns.
Users may need to compare temporal trends between two regions and compare spatial distributions over time~\cite{AndrienkoST2010}, which requires setting multiple volume spotlights and saving various map snapshots.
Besides, to analyze causalities~\cite{DBLP:journals/tvcg/DengWXBZXCW22} and cascading effects~\cite{DBLP:journals/tvcg/DengWLBZSXW22} of spatiotemporal phenomena, not only are automated extraction models required but also supplementary views with graph visualizations.

\section{Conclusion}
We leverage well-established volume visualization techniques to revisit the space-time cube, proposing a large-scale spatial time (ST) series visualization technique called VolumeSTCube.
In particular, we transform large-scale ST series into volumetric data that is subsequently visualized with volume and surface rendering.
To enhance users' ability to explore the visualization, we design user-friendly interactions oriented to spatial, temporal, and spatiotemporal patterns of ST series.
VolumeSTCube is evaluated with a computation experiment, a real-world case study, and a controlled within-subject user study.

\bibliographystyle{IEEEtran}
\bibliography{template}


 


\begin{IEEEbiography}[{\includegraphics[width=1in,height=1.25in,clip,keepaspectratio]{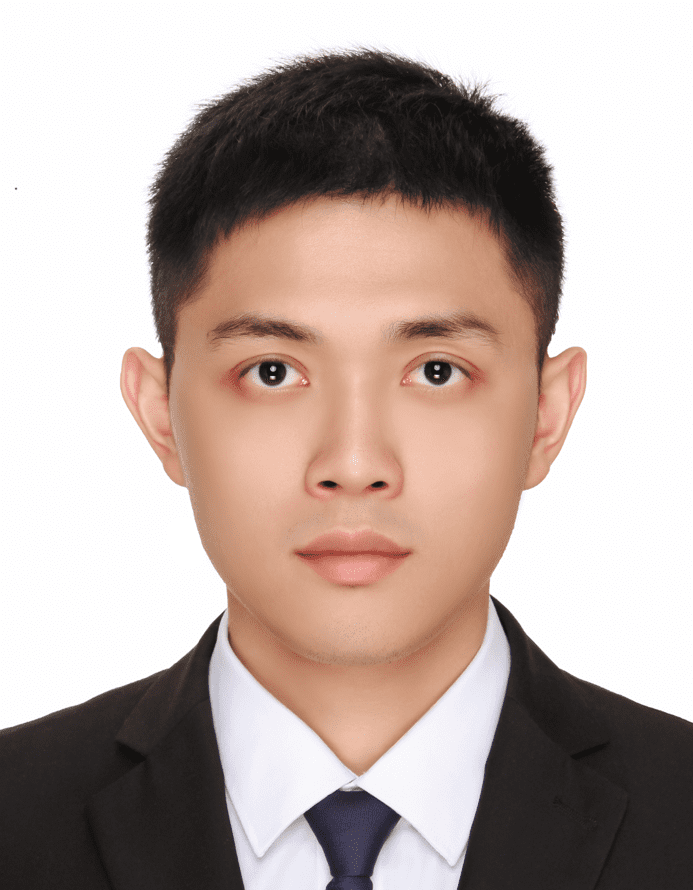}}]{Zikun Deng} is a tenure-track associate professor at School of Software Engineering, South China University of Technology. He received his Ph.D. degree in Computer Science from the State Key Lab of CAD\&CG, Zhejiang University in 2023. His research interests mainly include visual analytics, visualization, data mining, and their application in smart city, industry 4.0, and digital twins. He has published more than 10 papers in IEEE TVCG. For more information, please visit https://zkdeng.org.
\end{IEEEbiography}

\begin{IEEEbiography}[{\includegraphics[width=1in,height=1.25in,clip,keepaspectratio]{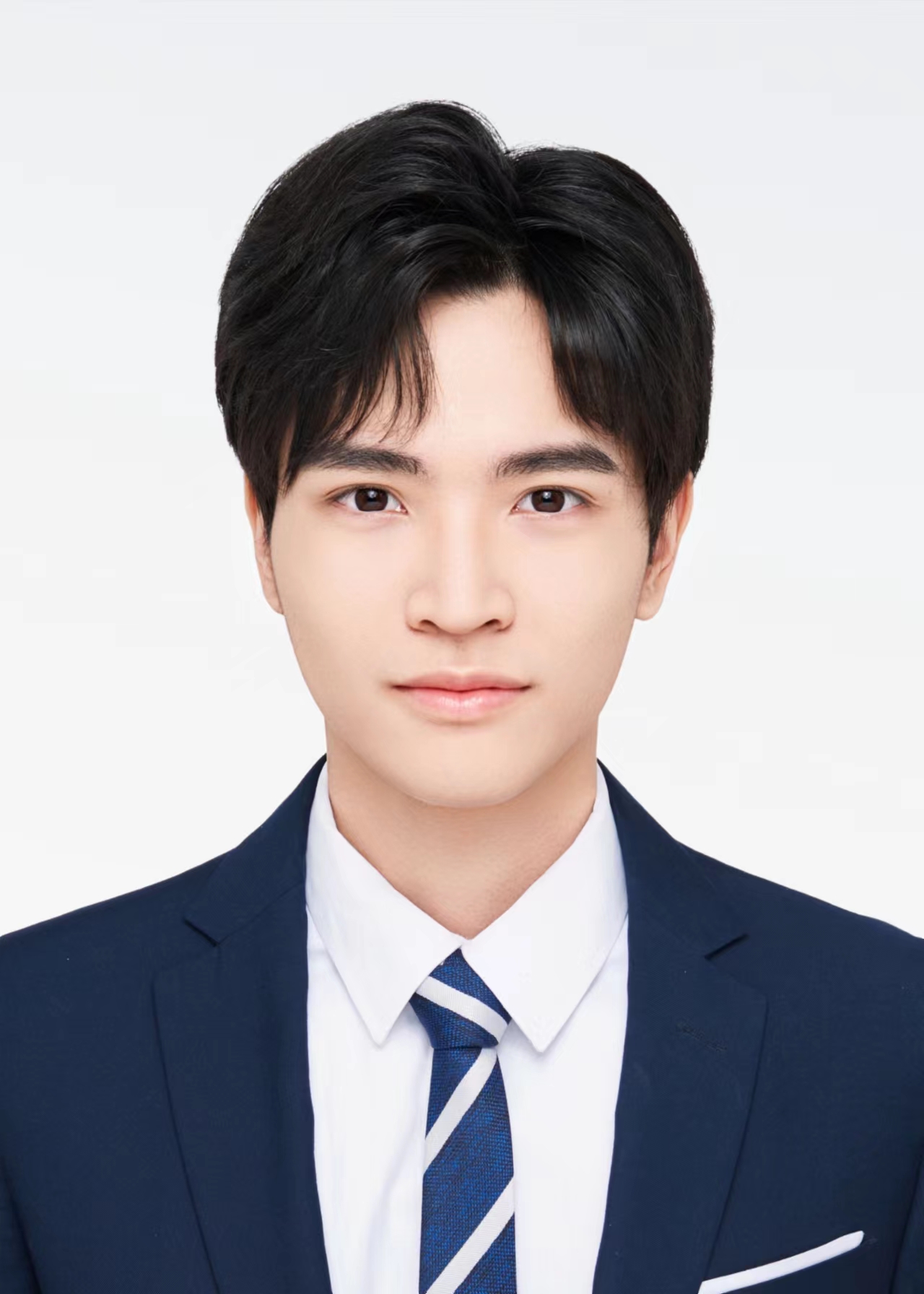}}]{Jiabao Huang} is an undergraduate student at the School of Software Engineering, South China University of Technology, and his research interests are data visualization, visual analytics, and computer graphics.
\end{IEEEbiography}

\begin{IEEEbiography}[{\includegraphics[width=1in,height=1.25in,clip,keepaspectratio]{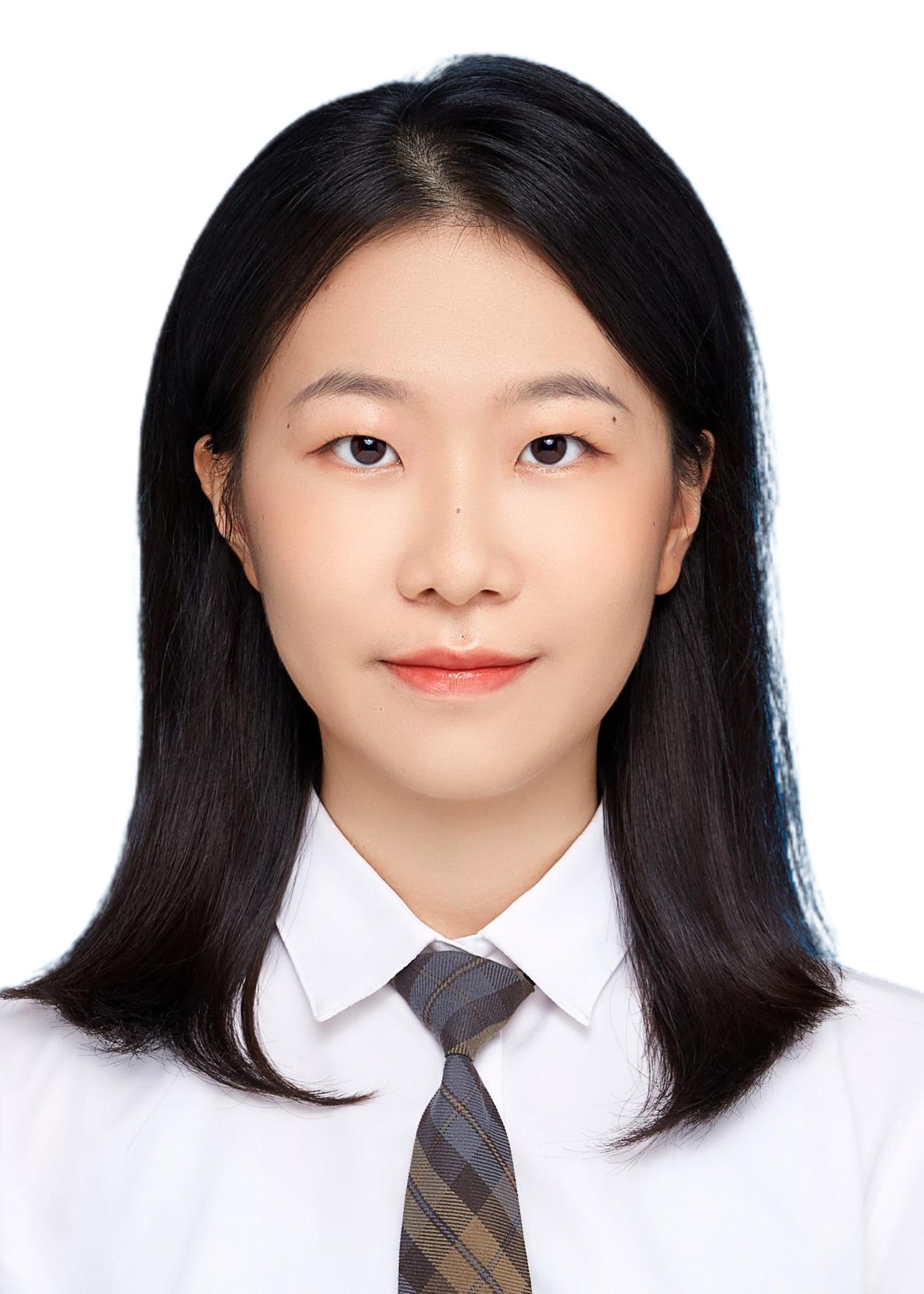}}]{Chenxi Ruan} is an undergraduate student at the School of Software Engineering, South China University of Technology, and her research interests are data visualization and visual analytics.
\end{IEEEbiography}

\begin{IEEEbiography}[{\includegraphics[width=1in,height=1.25in,clip,keepaspectratio]{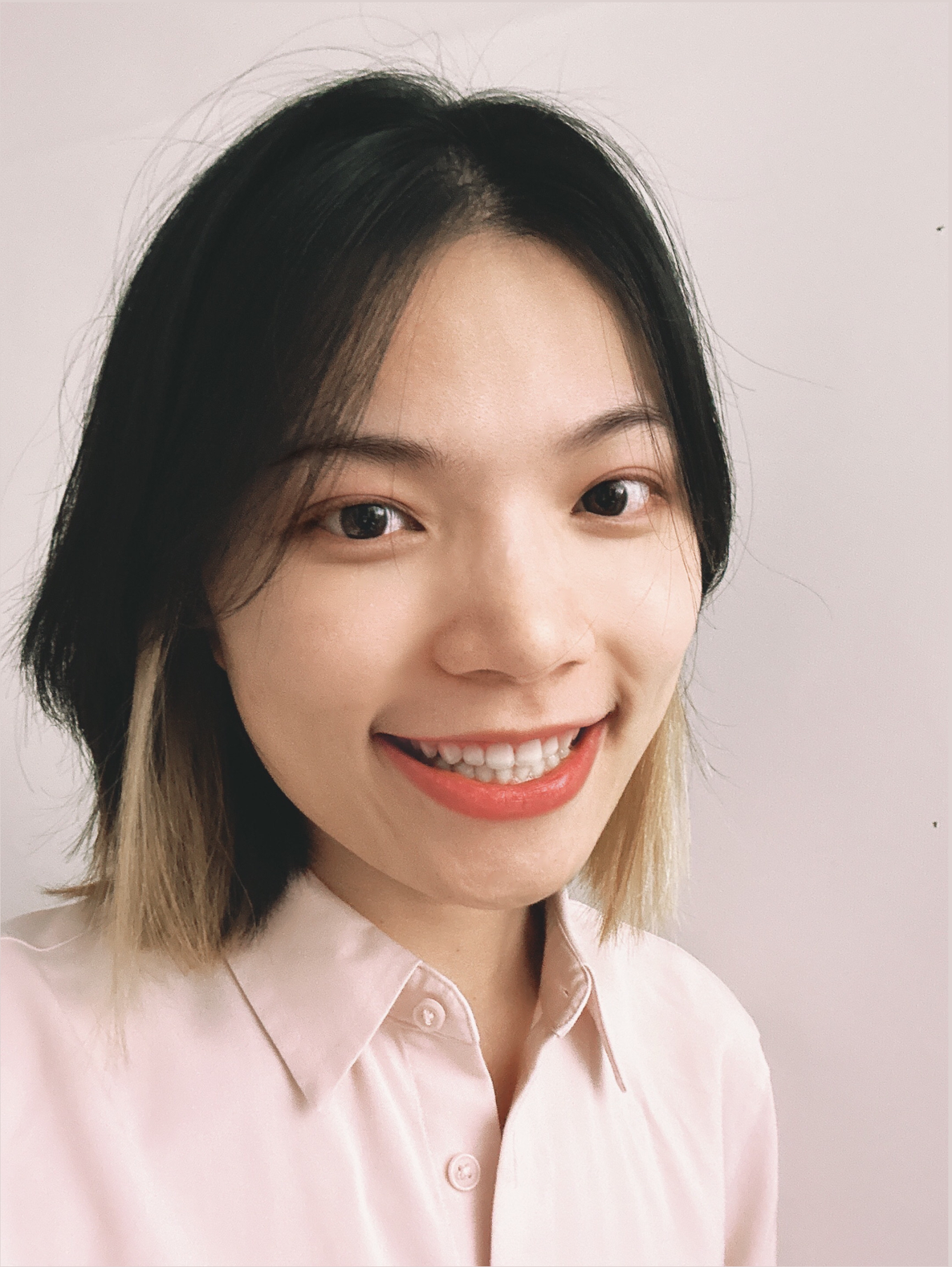}}]{Jialing Li} is a Ph.D. candidate at the Institute of Psychiatry, Psychology \& Neuroscience at King’s College London. Her main research interests are autism, social interaction, participatory research, and intervention development. She uses both quantitative and qualitative methods.
\end{IEEEbiography}

\begin{IEEEbiography}[{\includegraphics[width=1in,height=1.25in,clip,keepaspectratio]{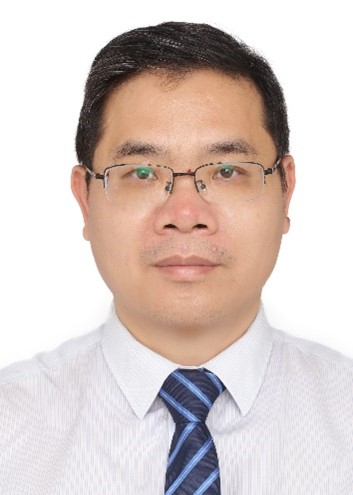}}]{Shaowu Gao} is currently the chief technical expert of the Greater Bay Area National Center of Technology Innovation. He received his Ph.D. degree from Shanghai Jiaotong University in 2004. He was a faculty in the Department of Mechanics at Wuhan University of Science and Technology. His main research interests are CAE, graphics rendering, and digital twins.
\end{IEEEbiography}

\begin{IEEEbiography}[{\includegraphics[width=1in,height=1.25in,clip,keepaspectratio]{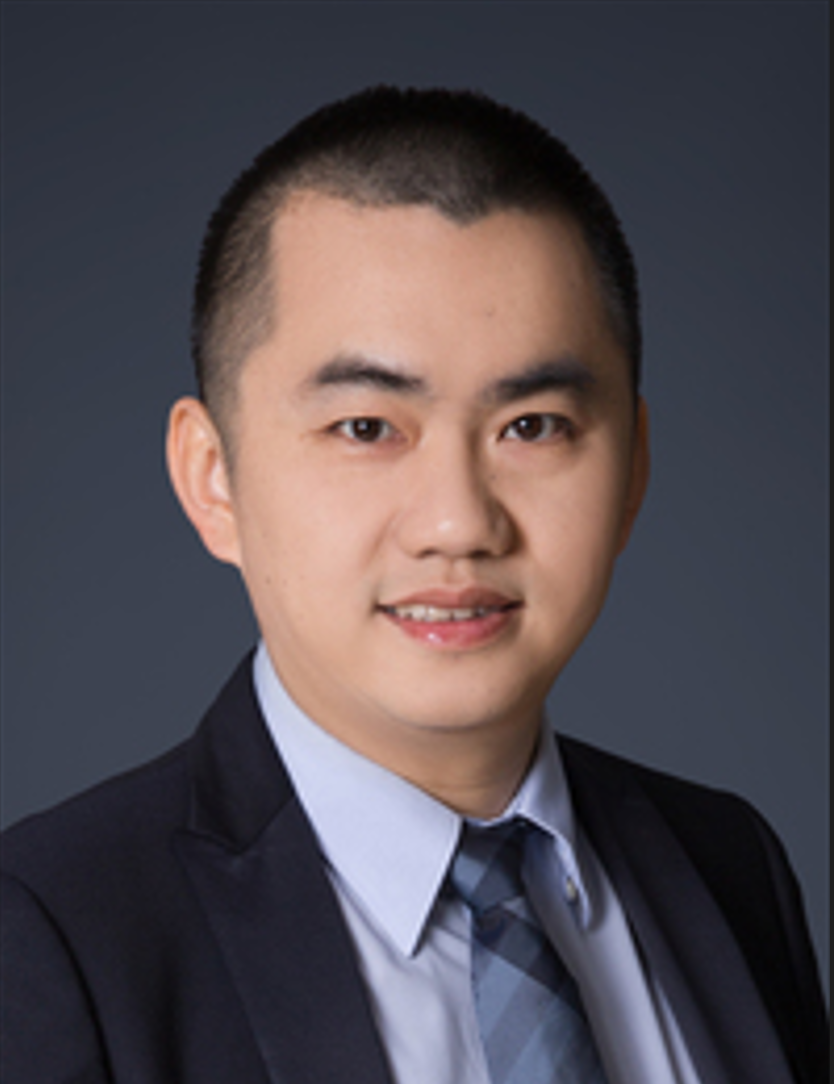}}]{Yi Cai} is the Dean of School of Software Engineering in South China University of Technology, the Director of The China Ministry of Education Key Laboratory of Big Data and Robotic Intelligence. He had received his PhD degree in the Chinese University of Hong Kong, and work as postdoctoral fellow in City University of Hong Kong. He has published more than 170 high quality papers in top journals and conferences such as IEEE TVCG, TKDE, TMM, AAAI, ACL, and ACM MM. He also has published 2 academic monographs and acts as the Chairman and Program Committee members of more than 20 prestigious international academic conferences.
\end{IEEEbiography}



\vfill

\end{document}